\documentclass[twocolumn]{emulateapj}
\usepackage{apjfonts}

\usepackage{graphicx}
\usepackage{mathrsfs}
\usepackage{natbib}
\bibpunct[,]{(}{)}{;}{a}{}{,}

\def\tff{{t_{\rm ff}}}
\def\rs0{{r_{\rm s0}}}
\newcommand{\ud}{\mathrm{d}}
\newcommand{\bomega}{\mbox{\boldmath$\omega$}} 
\newcommand{\bnabla}{\mbox{\boldmath$\nabla$}} 

\shorttitle{ACCRETION SHOCK WITH NUCLEAR DISSOCIATION}
\shortauthors{FERN\'ANDEZ \& THOMPSON}
\slugcomment{Accepted to ApJ}

\begin{document}

\title{Stability of a Spherical Accretion Shock with Nuclear Dissociation}
\author{Rodrigo Fern\'andez\altaffilmark{1} and Christopher Thompson\altaffilmark{2}}
\altaffiltext{1}{Department of Astronomy and Astrophysics, University of Toronto. 
Toronto,Ontario M5S 3H4, Canada.}
\altaffiltext{2}{Canadian Institute for Theoretical Astrophysics. 
Toronto, Ontario M5S 3H8, Canada.}

\begin{abstract}
We examine the stability of a standing shock wave within a spherical accretion 
flow onto a gravitating star, in the context of core-collapse supernova explosions. 
Our focus is on the effect of nuclear dissociation below the shock 
on the linear growth, and non-linear saturation, of non-radial oscillations of 
the shocked fluid. We combine 
two-dimensional, time-dependent hydrodynamic simulations 
using FLASH2.5 with a solution to the linear eigenvalue problem, and demonstrate
the consistency of the two approaches. 
Previous studies of this `Standing Accretion Shock Instability' (SASI) have focused either on
zero-energy accretion flows without nuclear dissociation, or made use of a detailed
finite-temperature nuclear equation of state and included strong neutrino heating.   Our main goal in this
and subsequent papers is to introduce  
equations of state of increasing complexity, in order to isolate the various competing effects.  
In this work we employ an ideal gas equation of state 
with a constant rate of nuclear dissociation below the shock, and do not include neutrino heating.
We find that a negative Bernoulli parameter below the shock 
significantly lowers the real frequency, growth rate, and saturation amplitude of the SASI.  A decrease in 
the adiabatic
index has similar effects.  The non-linear development of the instability is characterized by
an expansion of the shock driven by turbulent kinetic energy at nearly constant internal
energy.   Our results also provide further insight into
the instability mechanism:  the rate of growth of a particular 
mode is fastest when the radial advection time from the shock to the accretor 
overlaps with the period of a standing lateral sound wave.
The fastest-growing
mode can therefore be modified by nuclear dissociation.
\end{abstract}


\maketitle

\section{Introduction}

The explosion of a massive star involves the formation of a shock wave
surrounding its collapsed core.  The dynamics of this shock wave, and the
settling flow beneath it, are central to the explosion mechanism.  
A critical rate of neutrino heating below the shock will drive a purely spherical
explosion \citep{bethe85,burrows93}.  In practice this threshold 
may be reached only for low-mass cores (progenitor masses $\la~12 M_\odot$; 
\citealt{kitaura06}).   A successful spherical explosion does not develop in
more massive stars that form iron cores \citep{liebendoerfer01}.

For this reason, it has been generally believed that non-spherical effects are crucial
to the explosion mechanism, at least in the case of progenitor stars more massive
than $\sim~12\,M_\odot$. Several
related instabilities of this flow have been identified.  The first to be discovered
was a convective instability below the shock wave that is driven by neutrino heating
(\citealt{herant92}; see also \citealt{epstein79}).  There are also double-diffusive instabilities
associated with the transport of energy and lepton number 
at or below the neutrinosphere 
\citep{wilson88,bruenn95}.

The material which flows through the shock (predominantly iron, silicon, and oxygen) is
only weakly bound to the forming neutron star, but loses roughly half of its kinetic
energy to nuclear dissociation.  The Bernoulli parameter of the shocked
material with this dissociation energy subtracted becomes substantially negative.
Much of this dissociation energy can be restored if protons and neutrons recombine
into alpha particles, but this generally requires an outward expansion
of the shock from the radius of $\sim 100-150$ km at which it typically stalls.  
The successful expansion of the shock does not, however, require imparting positive energy
to the entire post-shock fluid:  buoyancy forces will drive a global finite-amplitude
instability in the presence of large-scale density inhomogeneities.  The stability 
analysis given in \citet{thompson00} shows that a non-spherical breakout of the shock is then possible.

Recent two-dimensional (2D) studies of the core collapse problem have revealed a persistent, dipolar
oscillation of the standing shock as an important degree of freedom in the collapse
problem.  In some circumstances, it has been shown that
low-order spherical harmonic perturbations (especially $\ell = 1$ and 2)
are linearly unstable while the flow below the shock is still laminar.  
A strong instability was uncovered in an axially symmetric
hydrodynamic simulation of a standing shock, in the case where the fluid has an ideal, polytropic
equation of state and the flow upstream of the shock is hypersonic \citep{BM03,BM06}.
This corresponds to the case where the settling flow below the shock has essentially
zero energy.  A linear stability analysis by \citet{F07},
which assumed the same ideal equation of state, revealed 
high frequency modes of
this instability to involve a feedback between ingoing entropy and vortex perturbations 
and an outgoing sound wave
which further perturbs the shock.  We provide further evidence for this interpretation
in our paper,
particularly for fundamental, low-order modes. 

Interest in the shock stability problem arises, in good part, from the 
prominence of the dipolar oscillation that is observed along the axis of 
symmetry in the most complete 2D core collapse simulations 
\citep{buras06a,buras06b,burrows07,marek07}.  The association of such an oscillation with
a linearly unstable 
mode (the Standing Accretion Shock Instability or SASI) 
should, however, be treated with caution 
when  both nuclear dissociation and neutrino heating 
effects are present.
A dipolar oscillation can
be excited by convective motions 
(\citealt{FT08b}, hereafter Paper II), and 
convection generally is present below the shock near the 
critical heating rate
for an explosion in the case of more massive progenitors.

Given the number of competing effects that enter into the full core collapse problem,
we have designed a set of hydrodynamic simulations of a standing accretion shock
which employ equations of state of increasing complexity.  In this paper, we use
an equation of state in which the fluid flow is nearly ideal, 
with a fixed specific energy removed from the flow immediately below the shock (so as
to mimic the most important effect of nuclear dissociation).  The cooling rate
is a strong increasing function of density and becomes significant only at some finite 
distance below the shock, which is freely adjustable. 
In the calculations reported in this paper, we do not include neutrino heating, with the
aim of separating the shock instability from the forcing effect of convection. 

We simulate the time evolution of the perturbed shock
using the FLASH2.5 code \citep{fryxell00}.   In the linear regime,
we demonstrate detailed agreement between 
hydrodynamic simulations and the solution to the differential
eigenvalue system of \citet{F07}.
The  linear growth rate is reduced dramatically as the ratio of the
dissociation energy to the gravitational binding energy at the shock is
increased, but we still find growing modes with several radial overtones.
A similar trend is observed when the adiabatic index is decreased. 
We also find that the nonlinear evolution of the shock is characterized by an expansion driven
by turbulent kinetic energy at nearly constant internal energy, with the kinetic energy
reaching $\lesssim 10\%$ of the time-averaged internal energy. 
The r.m.s. displacement of the shock from its equilibrium
position is also significantly reduced with increasing dissociation energy.
Regarding the linear
instability mechanism, we find that the most unstable, low-frequency
modes always lie in a region
where the period of standing lateral sound waves overlaps with the duration of the 
``advective-acoustic'' cycle.

The structure of this paper is the following. In \S2, we describe our initial conditions and
numerical setup. In \S3.1, we show that the linear stability analysis of \citet{F07} yields
the correct eigenfrequencies, by matching its predictions with values measured from time-dependent
simulations to high precision. In \S3.2 we use the linear stability to explore the effects of 
nuclear dissociation and adiabatic index in the linear phase. 
In \S4, we explore the saturated state of the SASI with time-dependent hydrodynamic simulations,
focusing on global energetics and time averaged properties. We address the linear instability
mechanism in \S5. Our work is summarized in \S6.

\section{Numerical Setup for Time-Dependent Simulations}\label{s:two}

We perform one- and two-dimensional, time-dependent hydrodynamic simulations with FLASH2.5 \citep{fryxell00}. 
This is a second-order Godunov-type Adaptive Mesh Refinement (AMR) code which implements the Piecewise Parabolic Method 
(PPM) of \citet{colella84}. The code 
allows for source terms in the equations of momentum and energy
conservation, which represent an external gravitational field and
spatially distributed heating and cooling,
as well as nuclear burning.

We employ a simplified model with no heating, similar to that used by \citet{BM06} and \citet{F07},
with an ideal gas equation of state of adiabatic index $\gamma$, the gravity of
a point mass $M$ located at the origin, and a constant mass accretion rate $\dot{M}$.  
We differ from these authors in the inclusion of an energy sink that represents the
(partial) dissociation of nuclei downstream of the shock.

In what follows, we describe our initial conditions and issues associated with a
time-dependent calculation.
 
\subsection{Initial Conditions}
\label{sec:initial_cond}

The initial state of our simulations is a steady and spherically symmetric flow, with a 
net rate of mass transfer $\dot M = 4\pi r^2 \rho |v_r|$ from the outer boundary to the 
center.  A standing shock is present at $r=r_\mathrm{s0}$, and the flow outside this shock
has a vanishing Bernoulli parameter
\begin{equation}
\label{eq:berc_def}
b \equiv \frac{1}{2}v_r^2 + \frac{\gamma}{(\gamma-1)}\frac{p}{\rho} - \frac{GM}{r} = 0
\qquad (r > r_{s0}),
\end{equation}
which ensures a vanishing energy flux through the shock.   In the above,
$v_r$, $p$ and $\rho$ are the radial velocity, pressure, 
and density of the fluid, $M = 1.3~M_{1.3}\,M_\odot$ is the gravitating mass, and $G$ is Newton's constant. 

The flow upstream of the shock has a finite pressure $p_1$ and Mach number
$\mathcal{M}_1 = |v_r|/c_{s1} = |v_r|(\gamma p_1/\rho_1)^{-1/2}$, which are
related to the upstream density by
$p_1 = (\rho_1/[\gamma\mathcal{M}_1^2]) (\dot{M}/[4\pi \rho_1 r_\mathrm{s0}^2])^2$.
There is no cooling or nuclear dissociation above the shock,
hence the conditions $\dot{M}=\textrm{constant}$, $b = 0$, and the equation of state completely determine
the initial flow for $r > r_\mathrm{s0}$.

The initial flow downstream of the shock is obtained by solving the time-independent 
continuity, Euler, and energy equations in spherical symmetry.  The effect of cooling is included as
a source term in the energy equation. 
The upstream and downstream solutions are connected through the Rankine-Hugoniot shock jump conditions 
that conserve the flux of mass, energy, and momentum across the shock (e.g., \citealt{landau}). 
We assume that 
nuclear dissociation is a nearly instantaneous process below the shock. When $r_\mathrm{s0}\sim 150$~km, the
gravitational binding energy per nucleon at the shock is 
$\sim 12 M_{1.3}(150\mathrm{ km}/r_\mathrm{s0})$~MeV, higher than the binding energy per nucleon 
of ${}^{56}$Fe, $\sim 8.8$~MeV (e.g., \citealt{audi03}).  
In our simplified model, the effect of dissociation is 
parameterized by a constant dissociation energy per unit mass $\varepsilon$.  
Specific internal energy
equal to $\varepsilon$ is removed from the fluid upon crossing the shock. To account for this process during 
initialization, the equation of energy conservation across the shock at $r = r_{s0}$ is modified to read
\begin{equation}
\label{eq:shock_epsilon}
\frac{1}{2}v_1^2 + \frac{\gamma}{\gamma-1}\frac{p_1}{\rho_1} = \frac{1}{2}v_2^2 +\frac{\gamma}{\gamma-1}\frac{p_2}{\rho_2} + \varepsilon.
\end{equation}
Here the subscripts $1$ and $2$ label the upstream and downstream flow variables.  
This yields a compression factor \citep{thompson00}
\begin{eqnarray}
\label{eq:kappa_phot}
\kappa \equiv \frac{\rho_2}{\rho_1} & = & 
(\gamma+1)\Bigg[ \left(\gamma + \mathcal{M}_1^{-2}\right) -\nonumber\\
 & &\left.\sqrt{ \left(1-\mathcal{M}_1^{-2}\right)^2 + (\gamma^2-1)\frac{2\varepsilon}{v_1^2}}\quad \right]^{-1},
\end{eqnarray}
which reduces to $\kappa \to (\gamma+1)/(\gamma-1)$ for $\mathcal{M}_1\to \infty$ and $\varepsilon = 0$.
Increasing $\varepsilon$ increases the compression factor
and decreases the post-shock Mach number.  For example, the complete dissociation of iron into nucleons 
costs an energy 
\begin{equation}
{\varepsilon\over v_\mathrm{ff}^2} \simeq 0.37\,M_{1.3}^{-1}\,\left({r_\mathrm{s0}\over 150~{\rm km}}\right),
\end{equation}
where $v_{\rm ff}^2 = 2GM/r_\mathrm{s0}$ is the free-fall speed.

The rate of release of internal energy per unit volume has the basic form 
\begin{equation}\label{eq:cool0}
\mathscr{L}_0 = Ap^\alpha \rho^{\beta-\alpha}.
\end{equation}
To make contact with previous calculations, we have adopted the
same parameterization as used by \citet{BM06}, $\alpha = 1.5$, $\beta = 2.5$.  This choice of exponents
is consistent with cooling by the capture of non-degenerate $e^+$ and $e^-$ on free nucleons 
when the pressure is dominated by relativistic particles.
($\mathscr{L}_0
\propto \rho T^6 \propto \rho p^{3/2}$). 
The rate of cooling due to the capture of degenerate, relativistic
electrons on protons is similarly a strong function of the pressure, 
$\mathscr{L} \propto Y_e\rho\mu_e^6 \sim \rho p^{3/2}$, where
$\mu_e \propto n_e^{1/3} \propto p^{1/4}$ is the electron chemical potential 
(the last relation assumes that the pressure is dominated by degenerate electrons).
For simplicity we use only a single cooling function in this work.  The constant normalization factor
$A$ determines the radius $r_*$ at which the flow stagnates, and thus the ratio $r_*/r_\mathrm{s0}$.
In assessing the effects of nuclear dissociation on the stability of the flow, we keep this
ratio constant and vary the normalization of the cooling function.

\begin{figure}
\includegraphics*[width=\columnwidth]{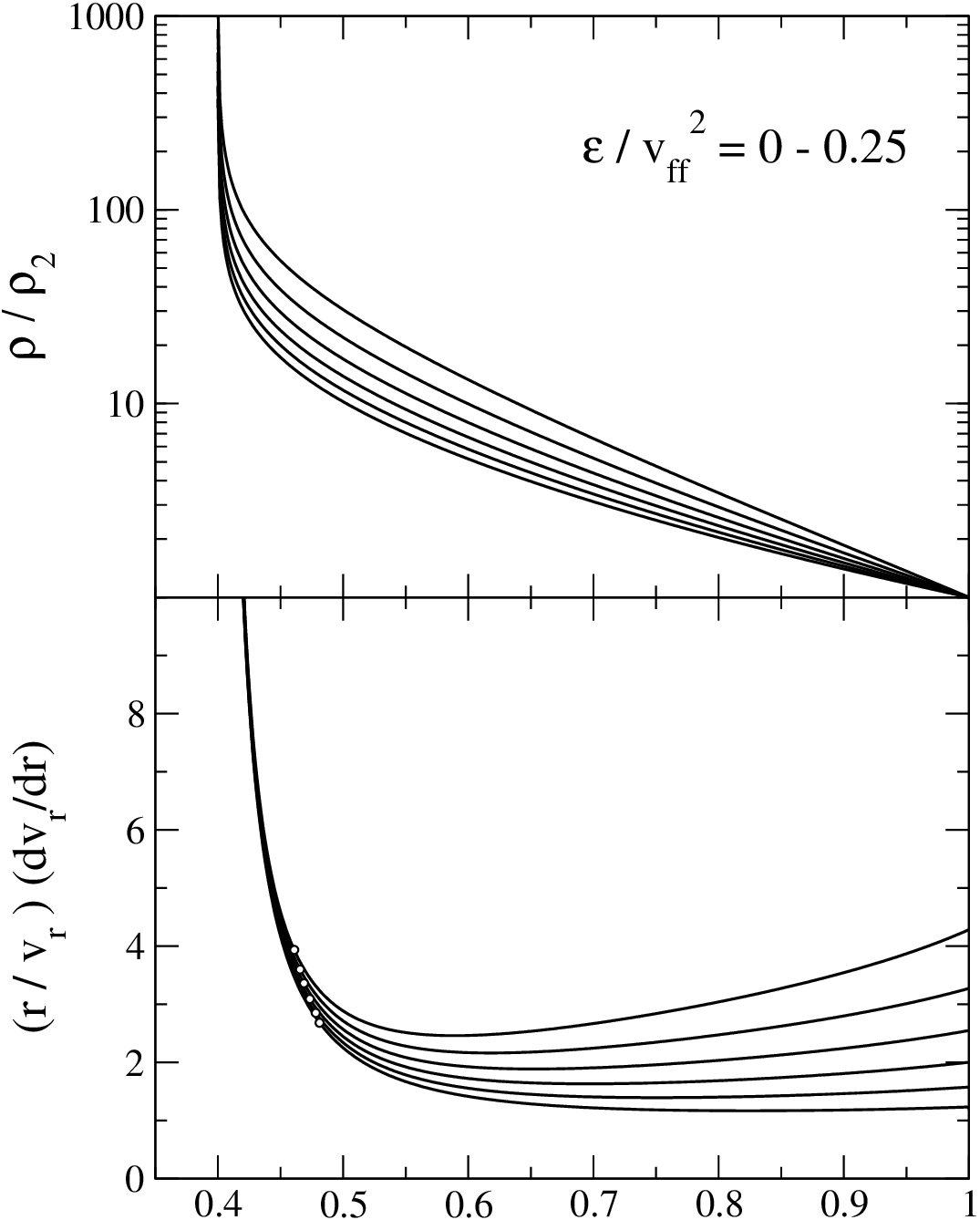}
\caption{Sample initial density profiles normalized to the postshock density $\rho_2$ (upper panel), and normalized radial velocity
gradient (lower panel), for dissociation
energies in the range $\varepsilon/v_{\rm ff}^2 = [0,0.25]$.
(Higher curves represent larger $\varepsilon$.)
Increasing the dissociation energy steepens the density profile as well as strengthening
the shock compression.  
The velocity gradient also 
becomes 
stronger just above the cooling layer (white circles denote the radius of maximum sound speed).
In this particular sequence, the normalization
of the cooling function $A$ is smaller by a factor of $127$ 
when $\varepsilon = 0.25v_{\rm ff}^2$ than when $\varepsilon = 0$.
}
\label{f:profiles_epsilon}
\end{figure}

Most of the energy loss is concentrated in a thin layer just outside the bound of the accretor at $r=r_*$.  
For the adopted scaling of $\mathscr{L}_0$ with $p$ and $\rho$, the ratio of cooling time to
flow time decreases with decreasing entropy.  The post-shock fluid then  undergoes runaway cooling 
in time-dependent simulations, and the shock collapses within a few sound crossing times. 
We have therefore implemented a cutoff in entropy in the cooling rate,
\begin{equation}
\label{eq:cooling_cutoff}
\mathscr{L} = \mathscr{L}_0\, \exp{[-(s/s_\mathrm{min})^2]},
\end{equation}
where $\mathscr{L}_0$ is the cooling function without cutoff (eq. [\ref{eq:cool0}]), 
$s = (\gamma-1)^{-1}\ln{(p/\rho^\gamma)}$ an entropy function,
and $s_\mathrm{min}$ the value of $s$ at $r=r_*$ that is obtained by evolving the
flow using the cooling function $\mathscr{L} = \mathscr{L}_0$.     
The result is that the accreted fluid accumulates in the first few computational 
cells outside $r_*$, with a minimal modification in the outer post-shock flow structure. 
The steady state flow so obtained is similar in structure to that calculated by \citet{HC92} 
for hyper-Eddington accretion onto a neutron star.  
For our choice of cooling exponents, and for fixed $r_*/r_\mathrm{s0}$ and $\gamma$, the main effects
of increasing $\varepsilon$ are an increase in the compression factor, a
smaller postshock Mach number, and a steepening of the density profile (see Fig.~\ref{f:profiles_epsilon}).
In going from $\varepsilon = 0$ to $\varepsilon = 0.25v_{\rm ff}^2$, the normalization of the cooling function 
is decreased by a factor of $127$ (keeping $r_*/\rs0$ and $\gamma=4/3$ fixed).

We normalize the radius to the initial shock radius $r_\mathrm{s0}$, all velocities to
the free-fall velocity at this radius, $v_\mathrm{ff}(r_{\rm s0}) = (2GM/r_\mathrm{s0})^{1/2}$, the
time to the free-fall time $r_{\rm s0}/v_{\rm ff}(r_{\rm s0})$, and 
the density to the initial upstream density $\rho_1$.  Numerical values appropriate for the
stalled shock phase of a core collapse are $r_\mathrm{s0} \sim 150$~km,  
$v_\mathrm{ff}(r_{\rm s0}) \sim 4.8\times 10^9 M_\mathrm{1.3}^{1/2}\,(r_{\rm s0}/150~{\rm km})^{-1/2}$~km s$^{-1}$, 
$t_\mathrm{ff} \equiv r_\mathrm{s0}/v_\mathrm{ff} \sim 3.1 \,M_\mathrm{1.3}^{-1/2}\,
(r_{\rm s0}/150~{\rm km})^{3/2}$~ms, and
$\rho_1 \sim 4.4\times 10^7 \dot{M}_\mathrm{0.3}\,M_\mathrm{1.3}^{-1/2}\,(r_{s0}/150~{\rm km})^{-3/2}$~g
cm$^{-3}$ (assuming a strong shock).   Here the mass accretion rate has been normalized to
$\dot M = 0.3\,\dot{M}_\mathrm{0.3}\,M_\odot$ s$^{-1}$.

\subsection{Time-Dependent Evolution}
\label{sec:time_dep}

Our computational domain employs spherical polar coordinates,
and the grid is uniformly spaced in both $r$ and $\theta$.  Two dimensional calculations make use of 
the full range of polar angles, $\theta = [0,\pi]$. 
We employ a baseline resolution 
$\Delta r_\mathrm{base} = r_\mathrm{s0}/320$ and 
$\Delta \theta_\mathrm{base} = \pi/192$.  
To better resolve the cooling layer,  one extra level of mesh refinement 
is added to all blocks satisfying $r < r_* + 0.1(r_\mathrm{s0}-r_*)$.  
To avoid excessively large gradients in this region and enhanced cooling due to
discreteness effects, we adjust the normalization $A$ of the cooling function
to satisfy
\begin{equation}
\bigg| \sum_i \mathscr{L}_{0,\,i}\, V_i \bigg| \simeq 0.995 
     \left( \frac{GM}{r_*} - \varepsilon \right) \dot{M},
\end{equation}   
where $V_i$ is the volume of the $i$-th cell, $\mathscr{L}_0$ is evaluated at the
lower end of the cell, and the sum is performed over the postshock domain. 
The numerical factor on the right-hand side is obtained empirically, and depends on
the radial resolution. This results 
in an initial Mach number $\sim 10^{-3} - 10^{-2}$ at the inner boundary.
The default FLASH2.5 Riemann solver is used, which we find can support high 
incident Mach numbers
$\mathcal{M}_1 \la 10^2$.  We have not witnessed the appearance of the odd-even decoupling instability 
at the shock \citep{quirk94}, which allows us to avoid using a hybrid Riemann solver.  
(We find that the hybrid solver in FLASH2.5 has problems for $\mathcal{M}_1\gtrsim 10$ in our setup.)

In all simulations reported in this paper, we use a reflecting inner boundary in the radial direction.
The flow at the outer radial boundary (situated at $2-4\,r_{\rm s0}$) is given by the upstream 
steady state solution. The angular boundaries at $\theta=0$ and $\theta=\pi$ are also reflecting 
in 2D simulations.

Our choice of a static and reflecting inner boundary is made for computational
simplicity.  In a real core collapse, the radius of the neutrinosphere decreases with time.
The effect of such a dynamic inner boundary on the linear growth of shock perturbations
has been examined by \citet{scheck08} in a semi-realistic collapse simulation.  They found that
it facilitated growth by increasing the strength of the velocity gradient above the neutrinosphere.
In addition, \citet{burrows06a} reported evidence for a transient SASI instability in the first 100 ms of
their full 2D collapse simulations.
The analysis of the SASI by \citet{scheck08}, like ours, used a reflecting inner boundary,
albeit one positioned inside the neutrinosphere.
Our results provide
strong evidence that linear growth depends on the structure of the flow between the
shock and the cooling layer, which suggests (but does not prove) that a more realistic mass distribution
below the cooling layer will not have a large impact on the growth of linear perturbations above 
the cooling layer. \citet{yamasaki07} find that a zero-gradient inner boundary condition
($\mathrm{d}\delta v_r/\mathrm{d}r = 0$) has a quantitative but not a qualitative effect
on the form of the linear eigenmodes.

At our baseline resolution, the accretion flow in 2D remains spherically symmetric for several tens or even
hundreds of dynamical times at the shock.  To excite specific modes of the flow,
we introduce an overdense shell in the flow upstream of the shock, with an angular dependence given
by a Legendre polynomial of a single order $\ell$.
In the absence of this perturbation, the flow still experiences a small startup error that is composed of two
spherically symmetric transients:  an outgoing
sound wave due to the finite (albeit small) Mach number at the inner reflecting boundary, and an ingoing
entropy wave due to the initial discontinuity at the shock (see, e.g., \citealt{leveque98}). 
Since these initial transients do not affect 
non-spherical 
modes, we made no attempt to suppress them by
increasing the 
numerical dissipation
(as is done by \citealt{BM03} and \citealt{BM06}), 
and instead use PPM in its default FLASH2.5 configuration
[see \citet{colella84,fryxell00} for details].
The spherical transients can be minimized (but never
completely eliminated) by locally increasing the resolution at the shock, 
and by simultaneously decreasing the innermost Mach number and increasing 
resolution at the inner boundary.

To account for nuclear dissociation at the shock, we use the \emph{fuel+ash} nuclear burning module 
in FLASH.  In the unmodified code, fluid labeled as
\emph{fuel} is completely transformed into \emph{ash} whenever certain bounds on density and temperature are exceeded.
A fixed energy $e_\mathrm{nuc}$ 
per unit mass is released and added to the internal energy of the fluid in an operator split way. 
In our implementation, the \emph{fuel} and \emph{ash} components both have the same ideal gas equation
of state, and are separated by a negative energy decrement 
$e_\mathrm{nuc} = -\varepsilon$.  
We aim to choose a threshold for ``burning'' (that is, nuclear dissociation) which maintains a composition of nearly
100\% fuel upstream of the shock, and 100\% ash downstream.  The composition of the fluid is divided this
way in the initial condition, with all fluid injected at later times through the 
outer radial boundary being entirely \emph{fuel}. 
Details of our implementation of nuclear burning 
and modifications to the original FLASH2.5 module are discussed in Appendix~\ref{s:burning}.

Potential numerical instabilities result from a combination of a high cooling rate and
a low flow Mach number in the cooling layer.  We find that the time integration is stable when
the cutoff (\ref{eq:cooling_cutoff}) is imposed on $\mathscr{L}$, because 
the cooling time is never smaller than the Courant time. 
(FLASH2.5 automatically restricts the timestep so as to satisfy the Courant-Friedrichs-Levy condition.)
Our implementation of nuclear dissociation also avoids introducing instabilities,
since the energy extracted from the flow in one time step is kept smaller than
the internal energy (Appendix~\ref{s:burning}).

The hydrodynamics module in FLASH has undergone extensive testing \citep{calder02}, and so we
focus our efforts on verifying our setup and on the interaction of the different physics modules 
with the 
hydrodynamic
solver. The most basic and complete test we can think of is the reproduction of the 
linear growth rates of the SASI.  The results are given in the next section.

\begin{figure*}
\begin{center}
\includegraphics*[width=0.9\textwidth]{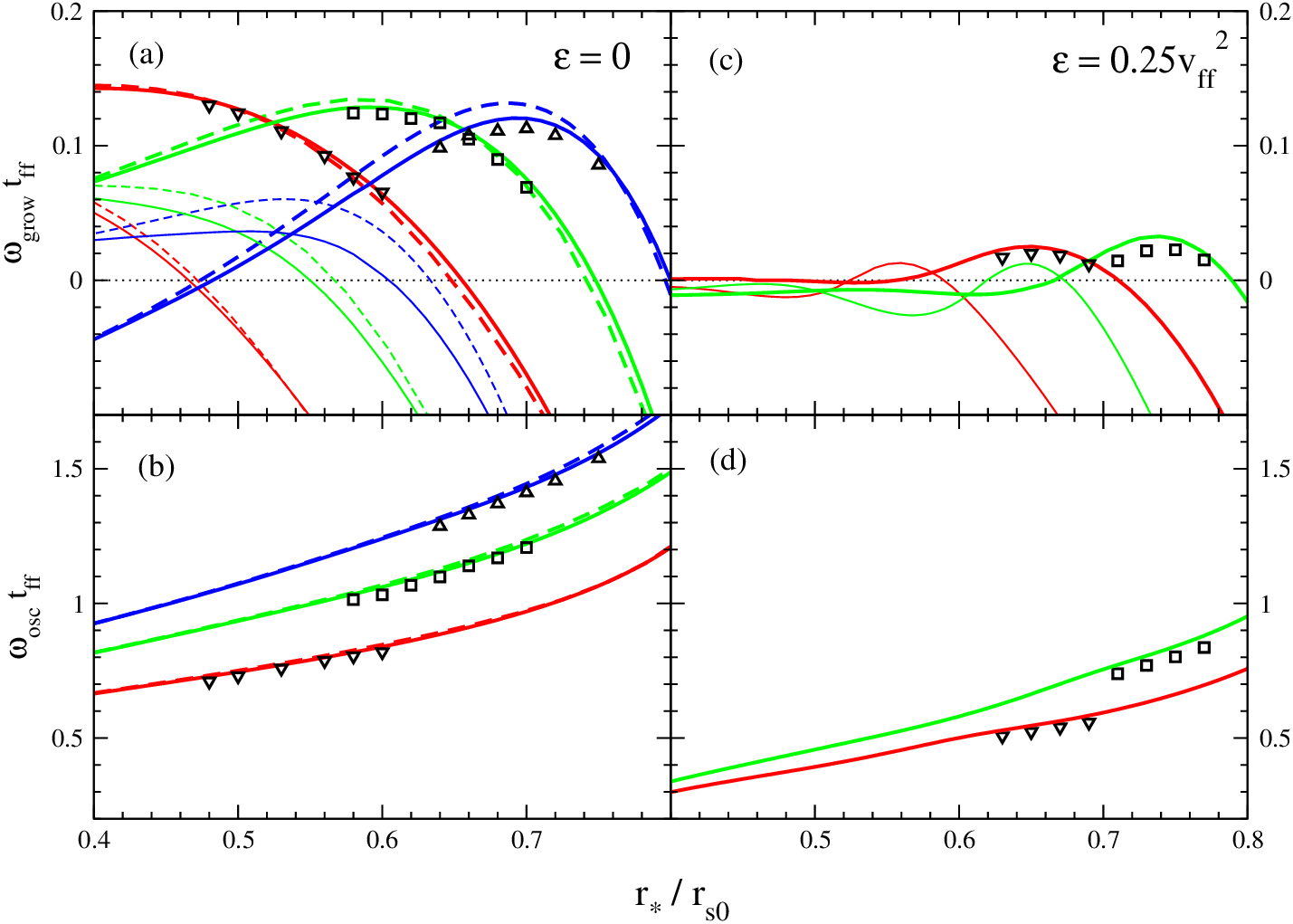}
\end{center}
\caption{Growth rates $\omega_\mathrm{grow}$ (upper panels) and oscillation frequencies $\omega_\mathrm{osc}$ (lower
panels) of the SASI, in units of inverse free-fall time at the shock, versus the
ratio of stellar radius to shock radius $r_*/r_\mathrm{s0}$.
In the left panels, the unperturbed flow has zero energy
flux
($\varepsilon=0$) and upstream Mach number
$\mathcal{M}_1 \simeq 87$.  In the right panels,
specific
internal energy $\varepsilon=0.25v_{\rm ff}^2$ is removed
from the flow just below the shock; and $\mathcal{M}_1=5$.  The lines give the solution to the eigenvalue
problem
of \citet{F07}
for different spherical harmonics:  $\ell=1$ (red), $\ell=2$ (green), and $\ell=3$ (blue),
with thick lines representing the fundamental mode and thin lines the first radial overtone.
The dashed lines show the effect of neglecting the cutoff in the cooling
function (eq. [\ref{eq:cooling_cutoff}]).  (See Appendix~\ref{sec:bndcnd_epsilon} for details.)
Symbols show the eigenfrequencies obtained from time-dependent hydrodynamic simulations:
for $\ell=1$ (down-triangles), $\ell=2$ (squares), and
$\ell=3$ (up-triangles). Uncertainties in the fitted parameters are smaller than the symbol size.
(See Appendix~\ref{sec:eigenmeasure} for details.)
}
\label{f:eigenfreq_e0.0}
\end{figure*}

\section{Linear Evolution}\label{s:three}

We now take a
look at the linear growth of perturbations of the accretion
flow below the shock.   We find a strong reduction in the growth rate when
nuclear dissociation is allowed to take place in the post-shock flow.
The time evolution of the perturbed flow is calculated with a direct
hydrodynamical simulation.  We have also modified the linear stability analysis of
\citet{F07} to include the effect of nuclear dissociation at the shock.   
The real and imaginary frequencies that are obtained by the two methods are compared.
The detailed agreement
that is obtained for a variety of modes provides a stringent test of both
the numerical simulation, and the solution to the eigenvalue problem.

\subsection{Comparison of Hydrodynamical Simulation and the Solution to the Eigenvalue Problem}

Our method for generating a perturbation and measuring its growth in a hydrodynamical simulation
is described in Appendix~\ref{sec:eigenmeasure};  and our
method for calculating the linear eigenmodes of the accretion flow is summarized in Appendix~\ref{sec:bndcnd_epsilon}.  
The three basic parameters of the flow are the ratio $r_*/r_{\rm s0}$ of cooling radius to shock radius,
the adiabatic index $\gamma$, and the 
dissociation energy $\varepsilon/v_{\rm ff}^2$.

It is possible to make a clean measurement of an individual SASI mode in a hydrodynamical simulation
when the fundamental is the only unstable mode (at a given $\ell$).  We have run simulations both in regions 
of parameter space where this condition is satisfied -- as predicted by linear stability analysis -- and 
where it is not.  The presence of unstable overtones can be gleaned from the time evolution of
a particular coefficient in the Legendre expansion of 
the shock radius, 
which deviates from a sinusoid of exponentially increasing amplitude.

First we reconsider the stability of the zero-energy accretion flow ($\varepsilon = 0$), and specialize
to an adiabatic index $\gamma = 4/3$.
In the parameter range relevant to core-collapse supernovae, we find that the most unstable modes are $\ell = 1$ 
and $\ell = 2$, in agreement with the work of \citet{BM06} and \citet{F07}.  The eigenvalue analysis
of \citet{F07}
correctly predicts the location of critical stability points of the fundamental and first radial overtone, for $\ell=\{0,1,2,3\}$. 
We have also solved the differential 
system of \citet{HC92}, finding that not only the critical stability points but also the 
growth rates do not agree with what we measure in our simulations (although results for $\ell=0$ are 
identical
to those of \citealt{F07}). 
In the remainder of this subsection, we refer to simulation results 
for which only the fundamental is unstable.

The left panels of Fig.~\ref{f:eigenfreq_e0.0} show growth rates and oscillation frequencies\footnote{
We have chosen to plot eigenfrequencies in units of the inverse free-fall time $\tff$, since this
is directly related to simulation time when both the stellar mass and the shock radius are held fixed.} 
for the modes $\ell=1,2,3$, in the case where the flow upstream of the shock has a high Mach number
$\mathcal{M}_1\approx 87$. 
There is very good agreement between the calculated eigenfrequencies and the output of the
hydrodynamic simulation:   $1-3\%$, $1-2\%$, and $1-2\%$ for the real frequencies of
the $\ell=1$, $2$, and $3$ modes, respectively, and  $2-4\%$, $2-8\%$, and 
$5-8\%$ for the growth rates.  Our method for estimating the uncertainty
in the measured mode frequencies is detailed in Appendix~\ref{sec:eigenmeasure}.  The most important systematic
errors arise from the discreteness of the mesh, the discrete summation involved 
in the Legendre projection, and the discrete time sampling.  The error bars are comparable to or smaller 
than the size of the symbols in Fig.~\ref{f:eigenfreq_e0.0}, 
typically $\delta \omega\, t_\mathrm{ff} \sim 5\times10^{-3}$.

The agreement between the two methods of calculating the growth rates becomes worse
at larger $\ell$.  We attribute this to the better sampling of the lower-$\ell$ modes by the grid,
which results in weaker numerical dissipation.  (The effects of numerical dissipation in PPM typically depend
on the ratio of 
cell width to wavelength:
\citealt{porter94}.)  

\begin{figure}
\includegraphics*[width=\columnwidth]{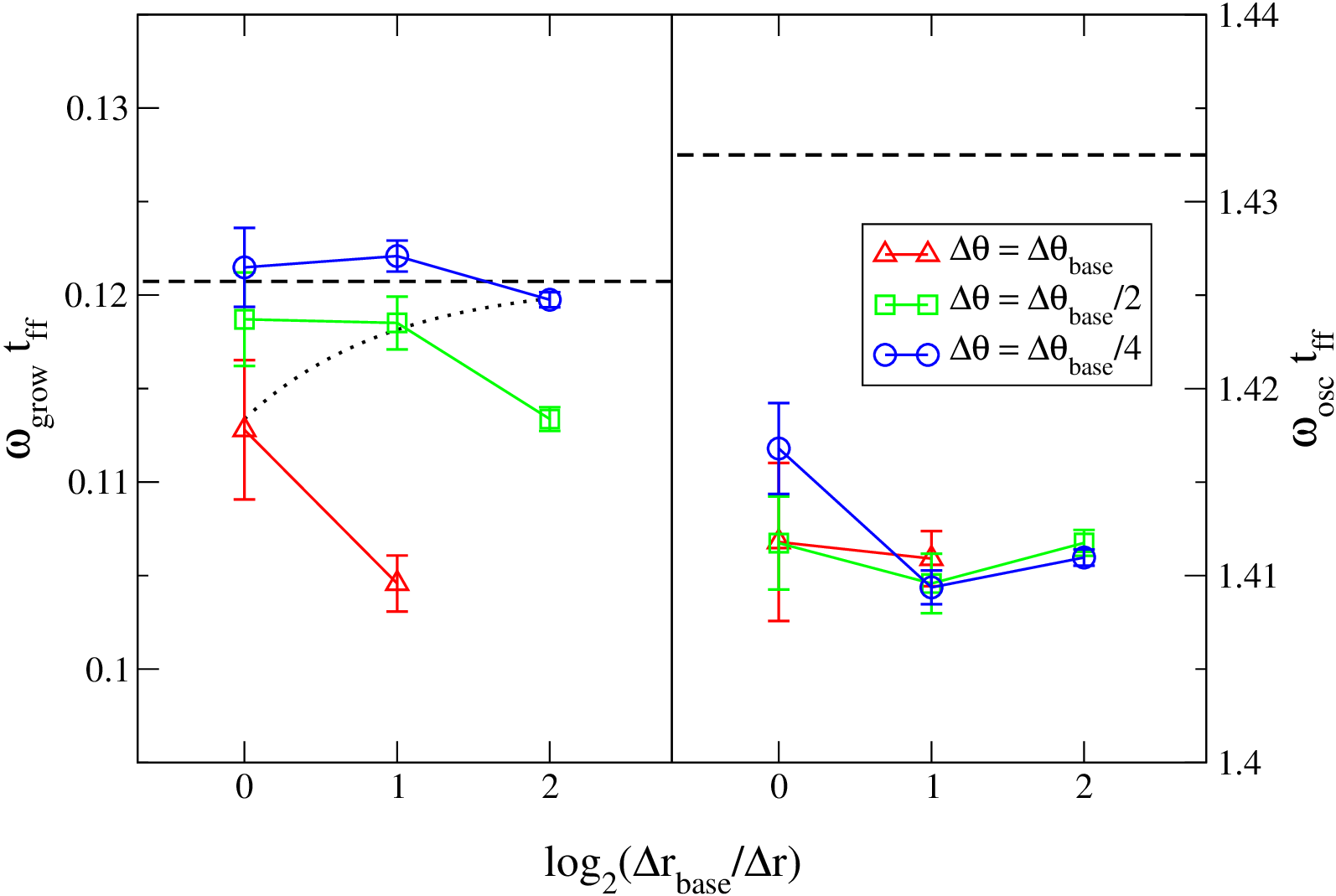}
\caption{Growth rates (left) and oscillation frequencies (right) as a function of resolution, 
for the $\ell = 3$ mode in a flow with $r_*/r_\mathrm{s0}=0.7$ (see panels a,b of Fig.~\ref{f:eigenfreq_e0.0}).
The dashed line in each panel shows the solution to the linear stability problem, and symbols represent
measurements from hydrodynamical simulations at different radial and angular resolutions.
See \S \ref{sec:time_dep} for a description of the grid spacing, and Appendix~\ref{sec:eigenmeasure}
for an explanation of the error bars. 
The dotted line
shows the best fit power law to the growth rate difference for a radial and angular
resolution increase: $\Delta \omega_\mathrm{grow} 
\propto (\Delta r \Delta \theta)^{-0.75\pm 0.10}$.}
\label{f:convergence_complete}
\end{figure}

\begin{figure*}
\includegraphics*[width=0.5\textwidth]{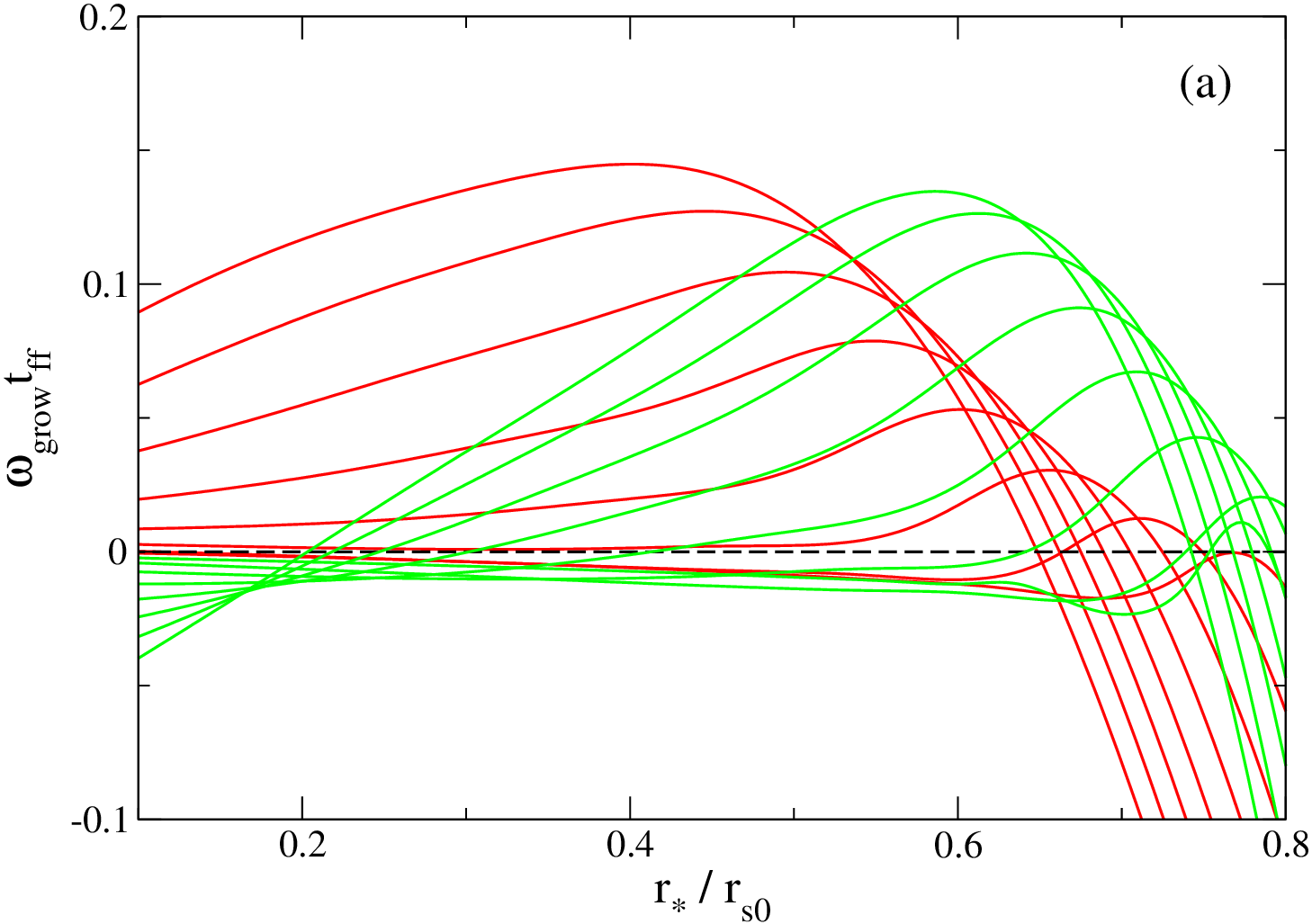}
\includegraphics*[width=0.5\textwidth]{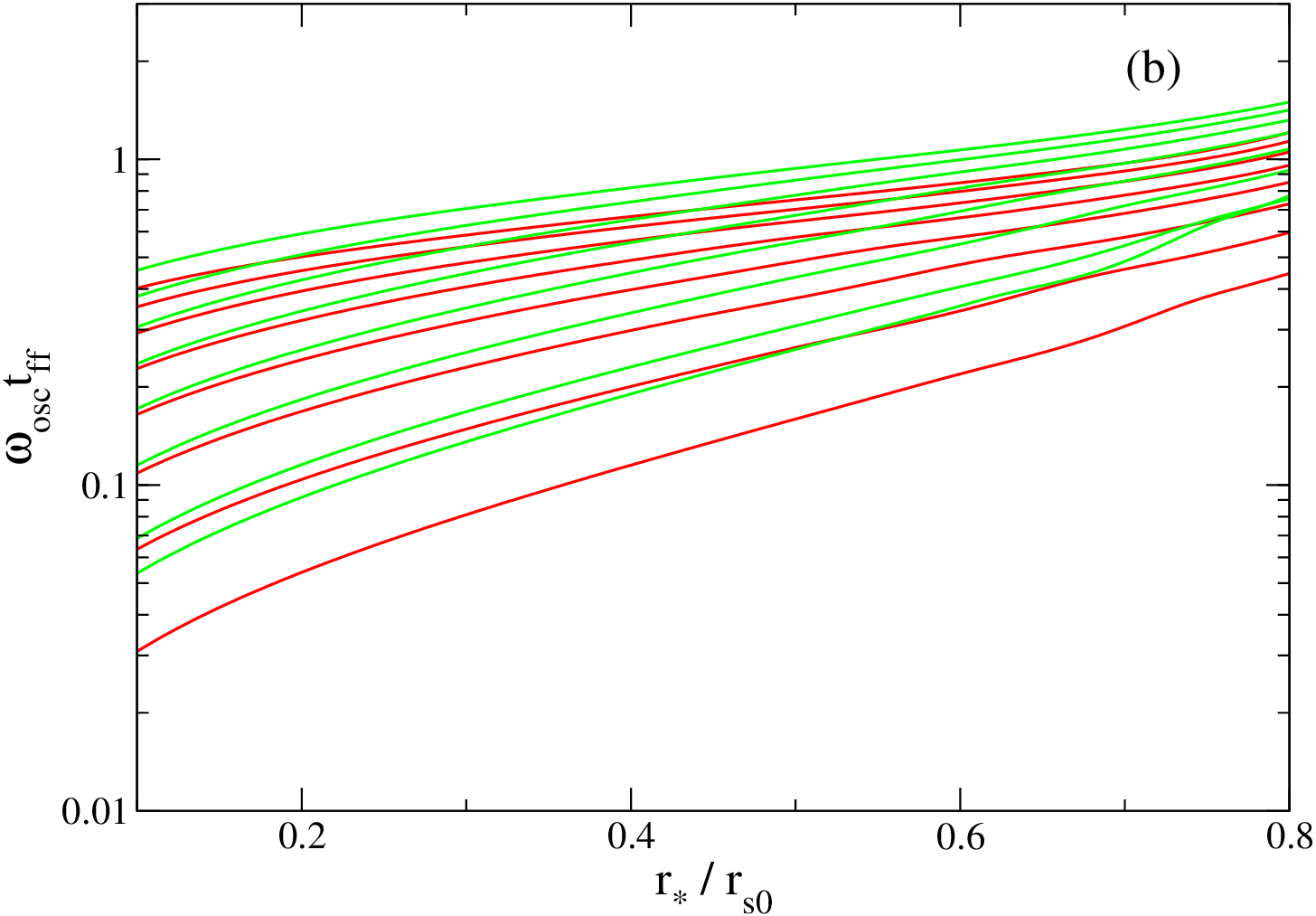}
\includegraphics*[width=0.5\textwidth]{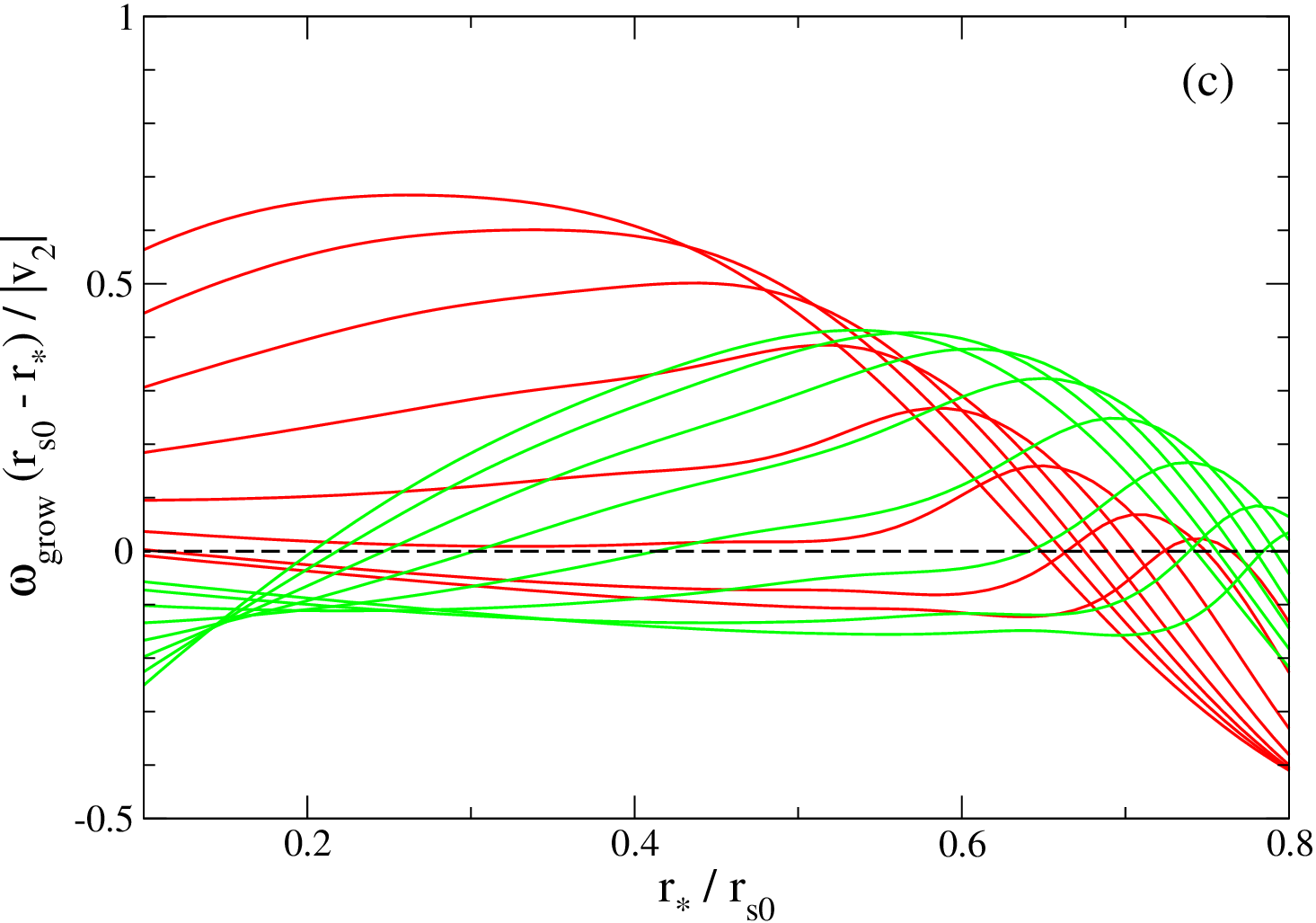}
\includegraphics*[width=0.5\textwidth]{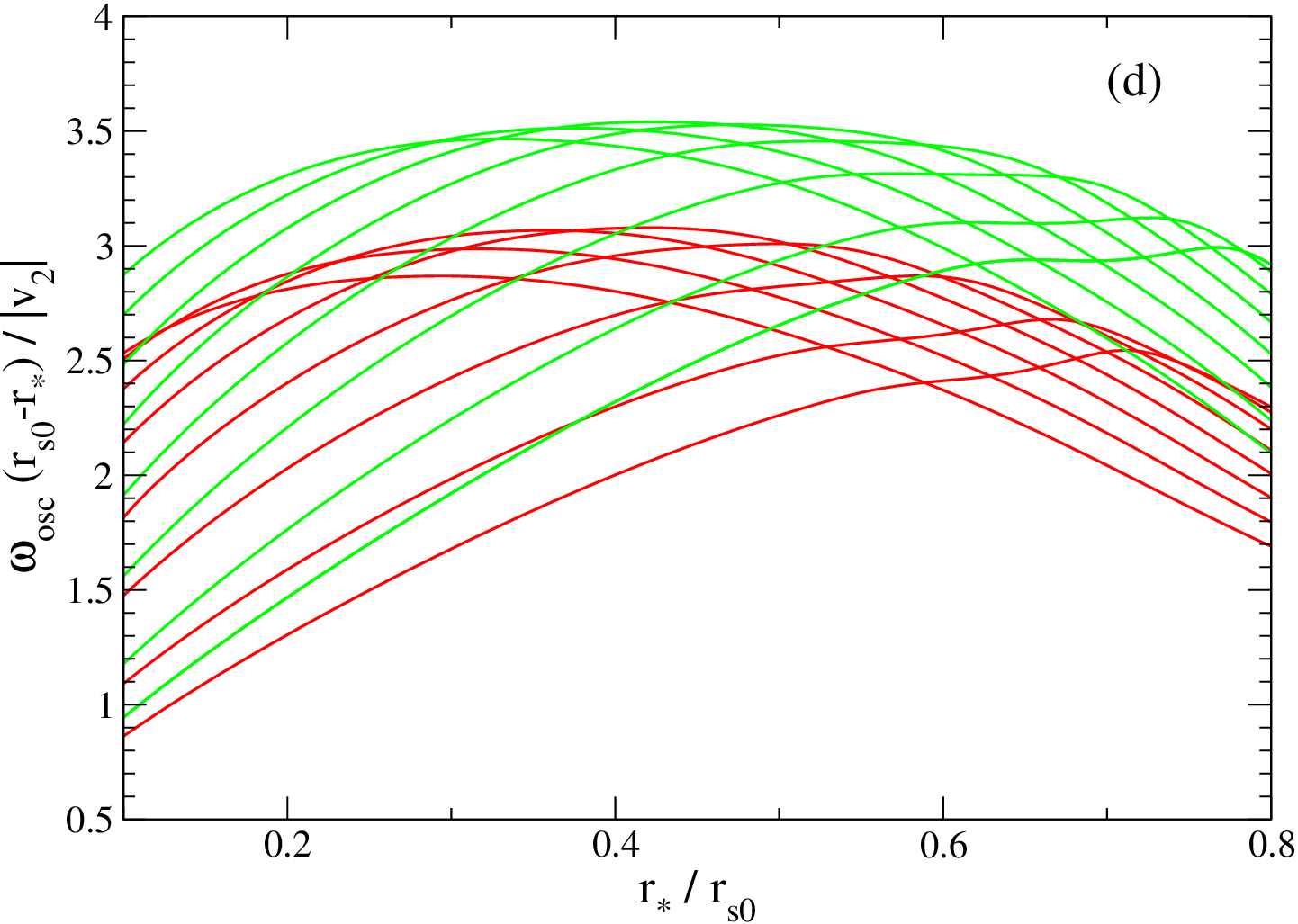}
\caption{Growth rates (left) and oscillation frequencies (right) of the 
fundamental
$\ell=1$ (red) and
$\ell=2$ (green) modes as a function of $r_*/r_\mathrm{s0}$, for dissociation energies
$\varepsilon/v_\mathrm{ff}^2 = 0,0.05,0.1,0.15,0.2,0.25,0.3$, and $0.33$.
Panels (a) and (b)
show eigenfrequencies in units of the inverse free-fall timescale $\tff$, with decreasing
curves corresponding to increasing $\varepsilon$. Panels (c) and (d) show the same curves, but in units
of the inverse of the time required to traverse the postshock cavity at the postshock speed, $(\rs0 - r_*)/|v_2|$. 
In panel (d), increasing dissociation makes the peak of the curves move to the right. 
Other parameters of the sequence are $\gamma=4/3$ and $\mathcal{M}_1\to \infty$. See text for description.}
\label{f:epsilon_fundamental}
\end{figure*}

\begin{figure}
\includegraphics*[width=\columnwidth]{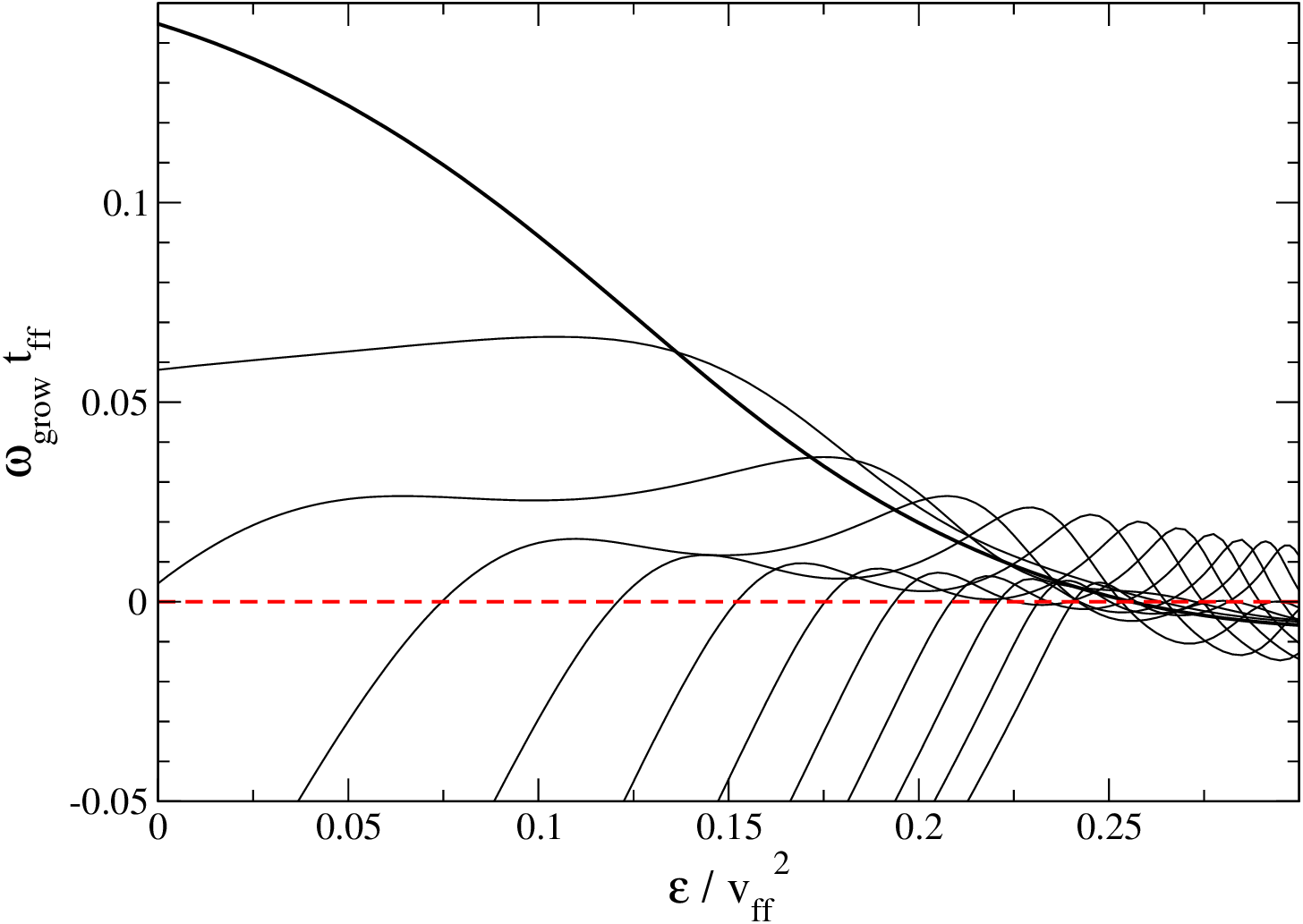}
\caption{Radial overtones of the $\ell=1$ mode as a function of nuclear dissociation energy $\varepsilon$.
The bold line denotes the fundamental mode, and the thin lines the first 11 overtones.
Other parameters are $r_*/r_\mathrm{s0}=0.4$,
$\gamma=4/3$, and $\mathcal{M}_1\to \infty$. Even though nuclear dissociation significantly decreases the
growth rate, there is always an unstable overtone.}
\label{f:epsilon_harmonic}
\end{figure}

\begin{figure}
\includegraphics*[width=\columnwidth]{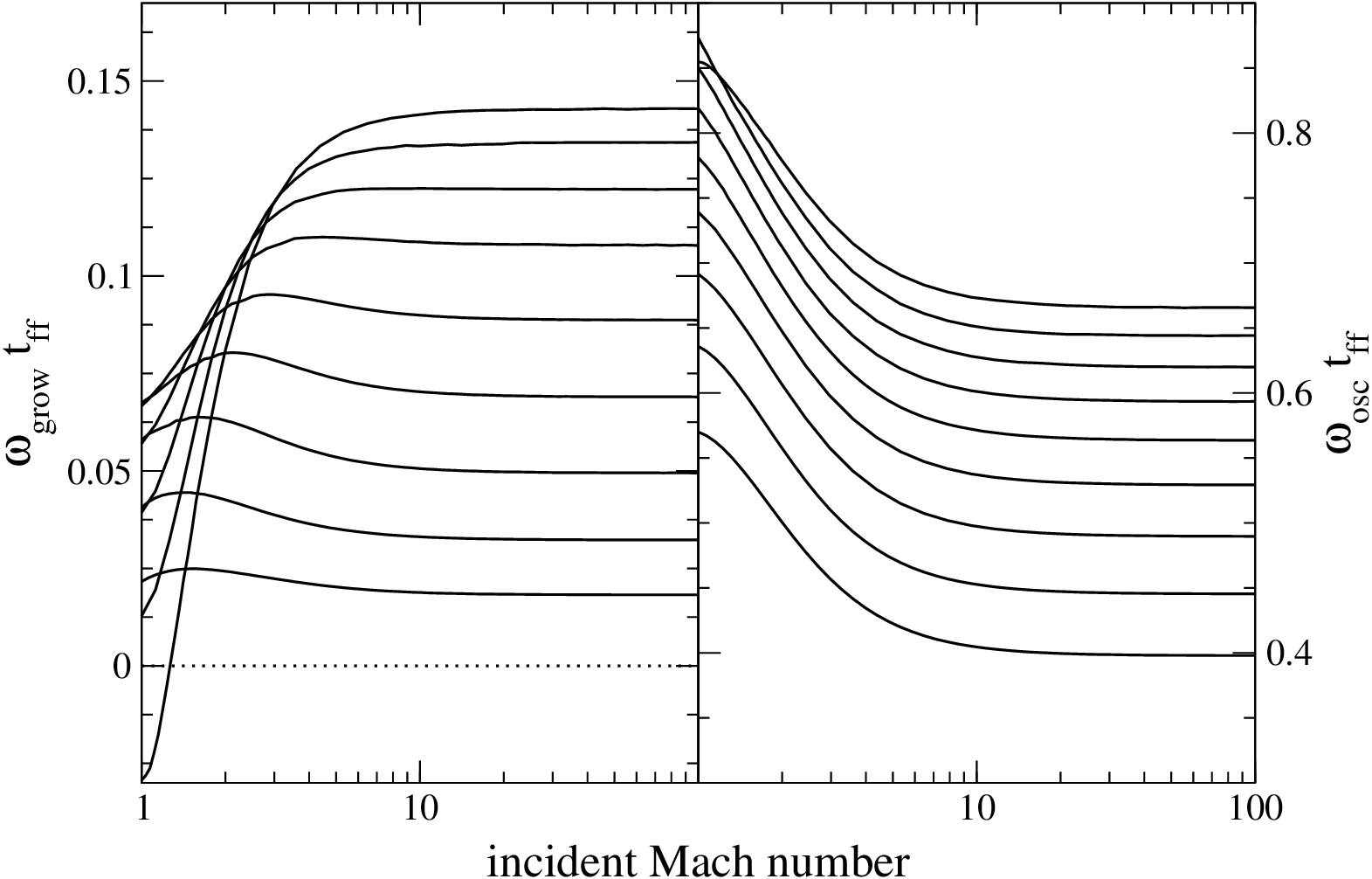}
\caption{Growth rates (left) and oscillation frequencies (right) of the 
fundamental
$\ell=1$ mode as a function of
incident Mach number, for $\varepsilon/v_{\rm ff}^2 = 0,0.025,0.05,0.075,0.1,0.125,0.15,0.175$, and $0.2$, with decreasing
curves for higher dissociation energy. 
Other parameters are $r_*/r_\mathrm{s0}=0.4$ and $\gamma=4/3$.
Eigenfrequencies are modified significantly when $\mathcal{M}_1 \lesssim 5$. The normalization of the cooling function
$A$ is bigger by a factor $4-5$ for $\mathcal{M}_1=2$ relative to $\mathcal{M}_1 = 100$ throughout the range in $\varepsilon$.}
\label{f:mach_eigenfreq}
\end{figure}

\begin{figure*}
\includegraphics*[width=0.5\textwidth]{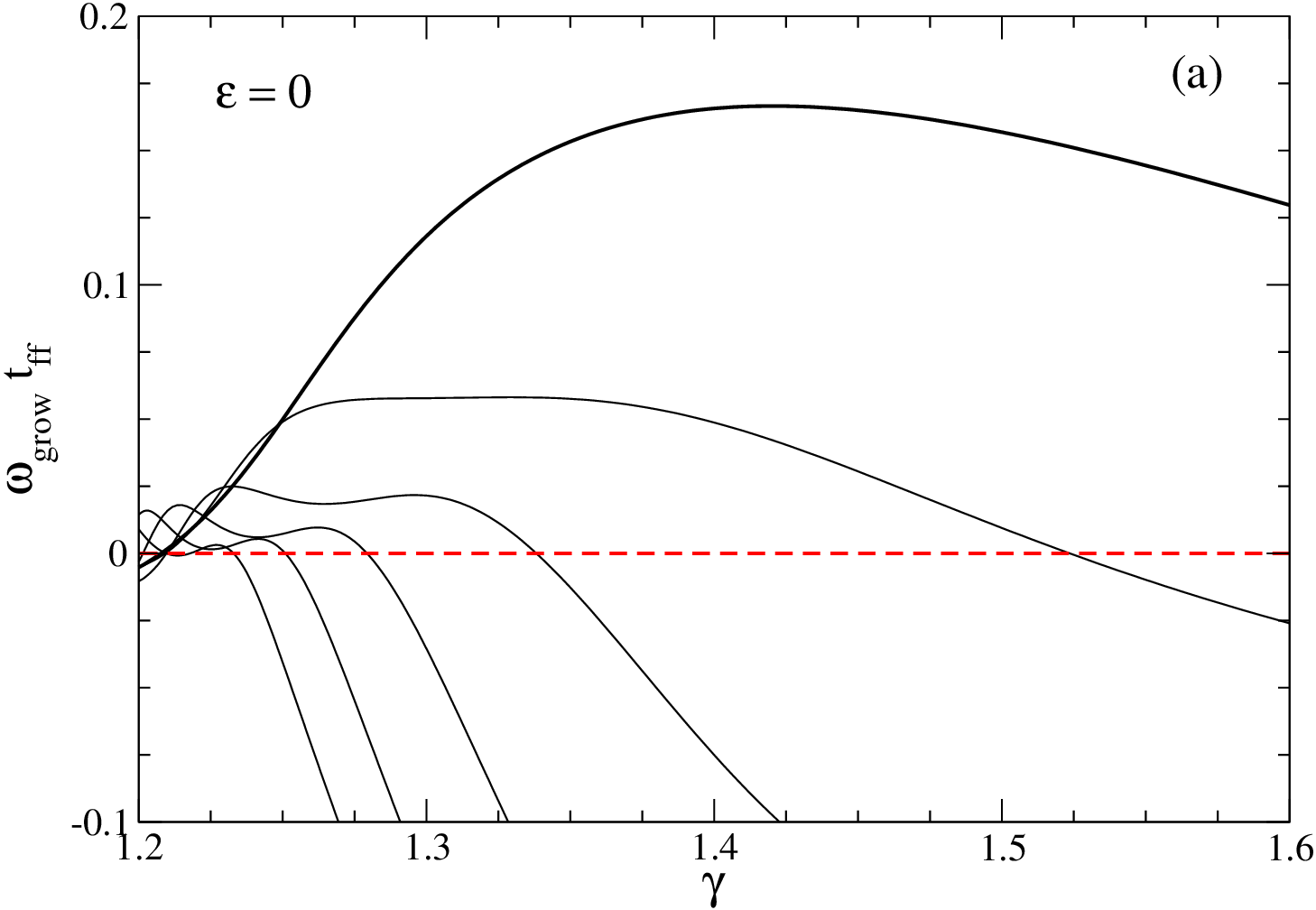}
\includegraphics*[width=0.5\textwidth]{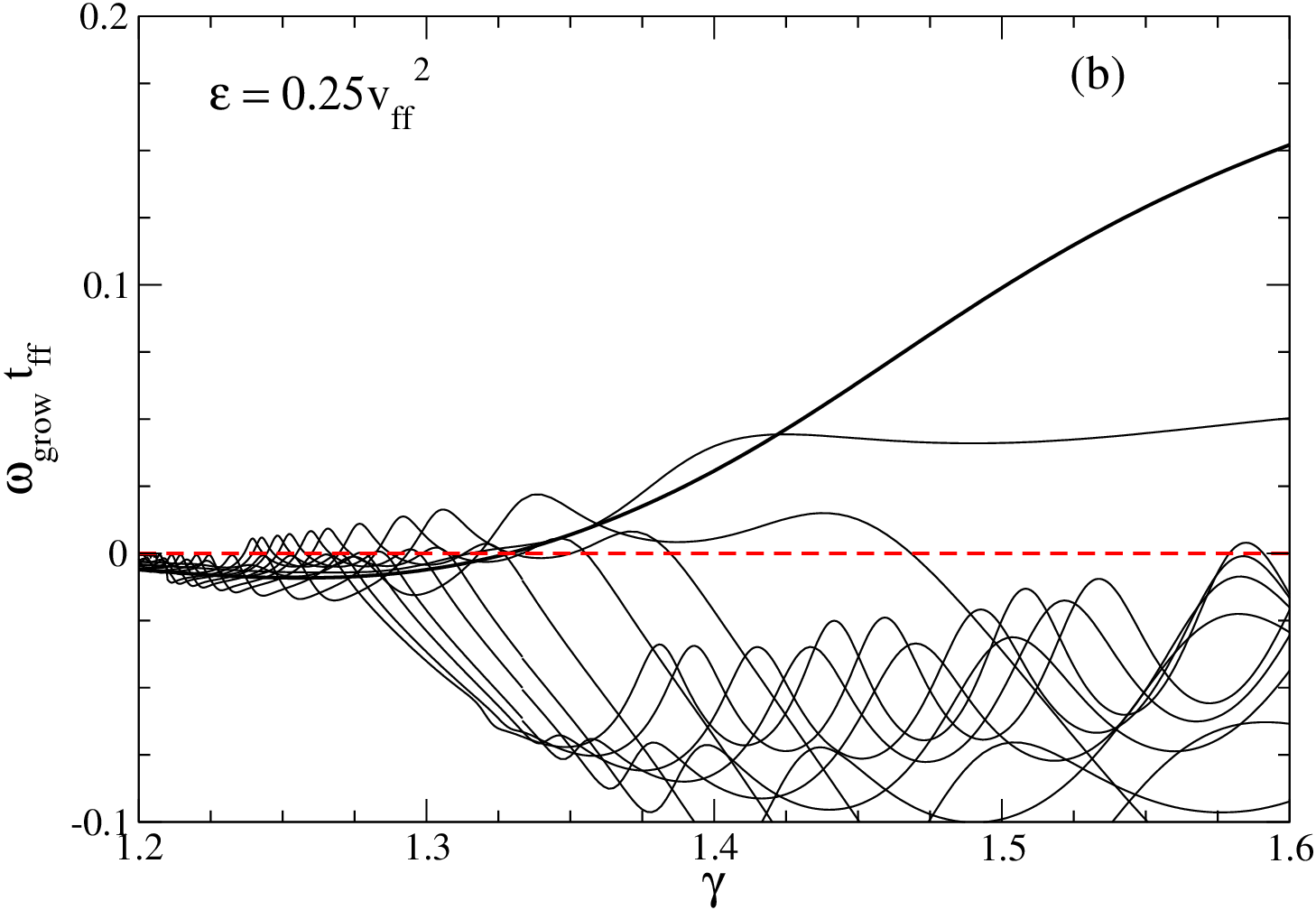}
\includegraphics*[width=0.5\textwidth]{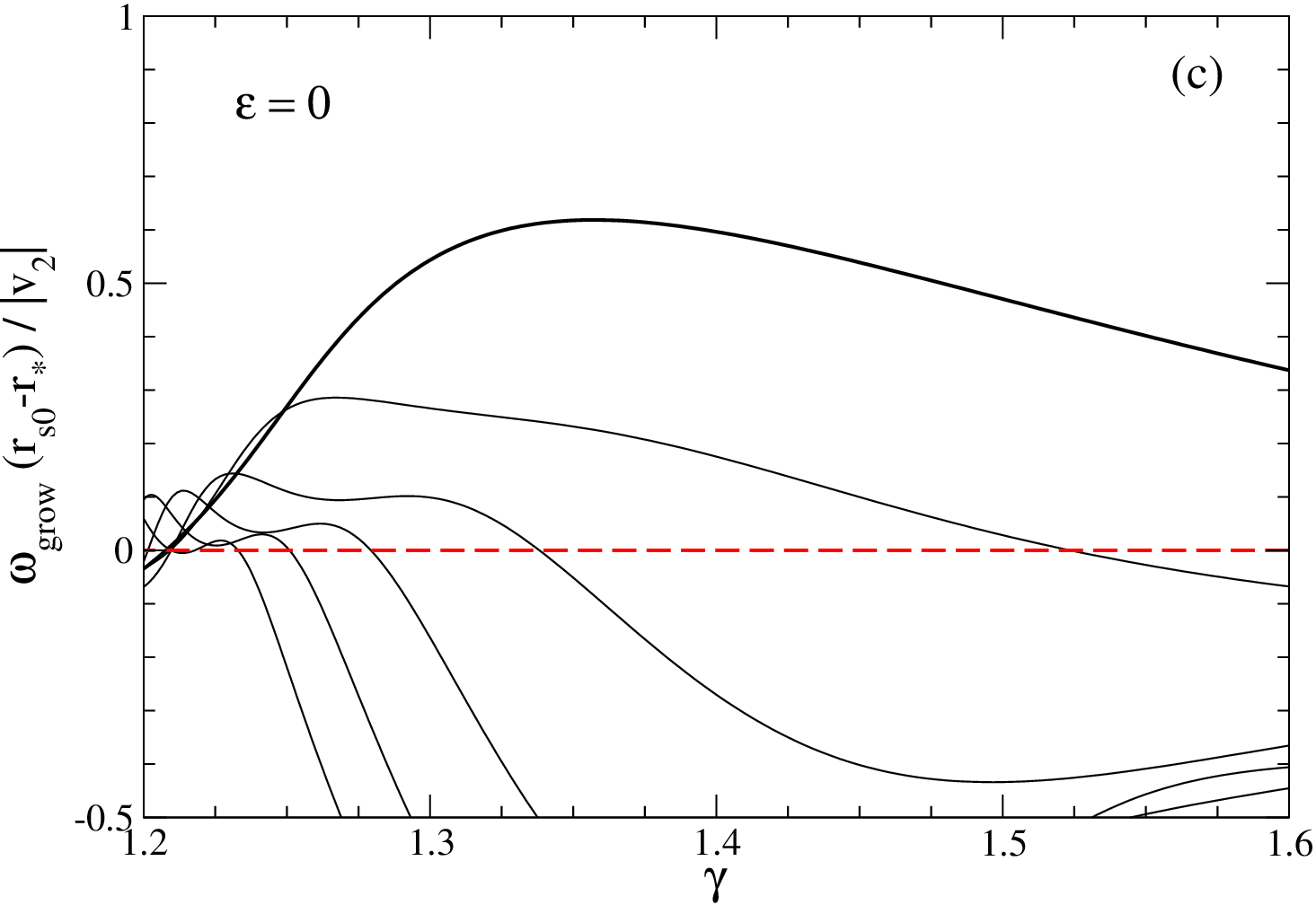}
\includegraphics*[width=0.5\textwidth]{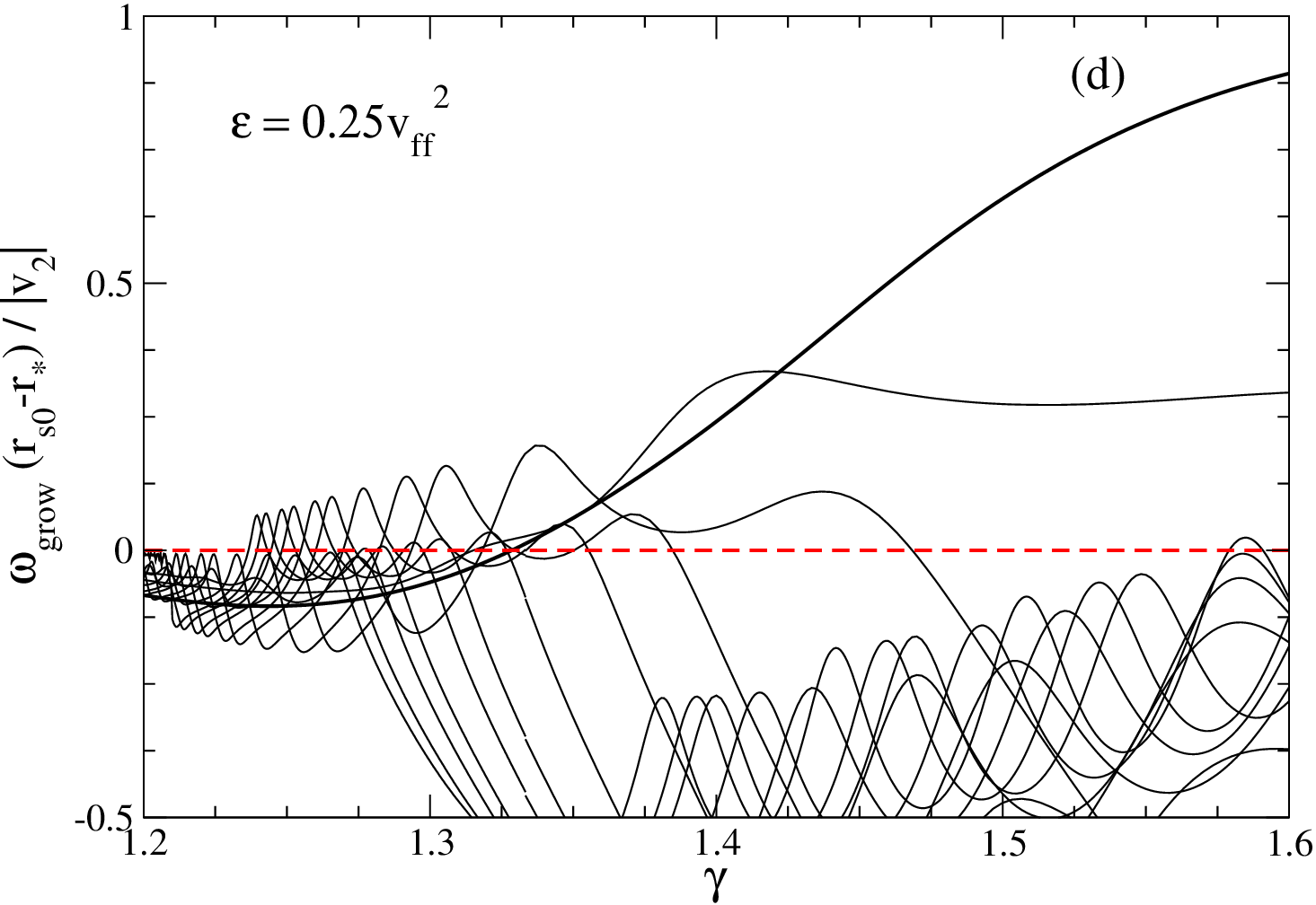}
\caption{Radial overtones of the $\ell=1$ mode as a function of adiabatic index $\gamma$, for a zero-energy
accretion flow [$\varepsilon=0$, panels (a) and (c)] and a flow with internal energy $\varepsilon=0.25v_\mathrm{ff}^2$
removed below the shock [panels (b) and (d)]. The upper row shows results in units of the inverse 
free-fall time $\tff$, whereas the lower row shows eigenfrequencies in units of the inverse of the time required
to traverse the shock cavity at the postshock velocity, $(\rs0 - r_*)/|v_2|$. The bold line denotes the fundamental
mode, the thin lines the first 6 (a,c) and 11 (b,d) overtones.
Growth rates generally decrease toward smaller $\gamma$, but there is always an 
unstable overtone. This qualitative result is independent of the size of the density jump at the shock.
Other parameters are $r_*/r_\mathrm{s0}=0.4$, $\varepsilon=0$, and $\mathcal{M}_1\to \infty$.}
\label{f:gamma_harmonic}
\end{figure*}

When the dissociation energy is increased to $\varepsilon = 0.25 v_{\rm ff}^2$, the
agreement between the two calculational methods is reduced a bit, to
$4-5\%$ for the oscillation
frequencies for both $\ell=1$ and $\ell=2$, and $18-30\%$ and $20-42\%$ for the growth 
rates, 
respectively.\footnote{The absolute value of the discrepancy between the two sets of growth rates
is similar to  that found for $\varepsilon=0$, namely 
$(10^{-3}-10^{-2})t_\mathrm{ff}^{-1}$.}
A more extensive exploration of the influence of finite $\varepsilon$ and a reduced upstream Mach number
on the mode frequencies is made in the next subsection.
For now, we emphasize the good agreement between analytics and numerics.

We have also checked the effects of resolution. 
Fig.~\ref{f:convergence_complete} shows results for
$r_*/r_\mathrm{s0} = 0.7$ and $\ell=3$ 
[see panels (a) and (b) of Fig.~\ref{f:eigenfreq_e0.0}].
The growth rates show the expected behavior, with increasing agreement between the two methods
with increasing radial and angular resolution.   We observe a greater sensitivity to the angular 
width $\Delta \theta$ of the grid cells than to their radial width.  The real frequencies show a 
much weaker trend. In sum, a four-fold decrease in $\Delta r$ and $\Delta \theta$ improves the 
agreement in the two sets of growth rates from $7\%$ to $0.8\%$, while the disagreement in the
real frequencies stays nearly constant within uncertainties at $1.5\%$.

\subsection{Dependence of Eigenfrequencies on Nuclear Dissociation and Adiabatic Index}

We now turn to the effect of nuclear dissociation on the 
oscillation frequencies and growth rates.  
As is observed in the right
panels of Fig.~\ref{f:eigenfreq_e0.0},  there is a substantial
reduction in growth rate when the dissociation
energy is raised to $\varepsilon=0.25v_\mathrm{ff}^2$ (and the Mach number upstream of the shock
is reduced to $\mathcal{M}_1 = 5$).    
One also observes a significant drop in the real frequency.
This effect is further illustrated in Figs.~\ref{f:epsilon_fundamental}a,b for the fundamental $\ell=1$
and $\ell=2$ modes, as the rate of nuclear dissociation is gradually increased.

Both trends can be partially explained in terms of an increase in the advection time below the
shock.  A finite nuclear dissociation energy at the shock increases the compression
factor $\kappa$ (eq.~\ref{eq:kappa_phot}) and decreases the post-shock Mach number 
\begin{equation}\label{eq:m2}
\mathcal{M}_2 =  \left[\kappa\left(\gamma+{1\over \mathcal{M}_1^2}\right)-\gamma\right]^{-1/2}.
\end{equation}
The advection time scales as $r_{\rm s0}/v_2 \propto \kappa$, whereas the lateral sound-travel
time has a weak dependence on $\kappa$.  As a result, the
advection time increases relative to the lateral sound travel time.
The growth rate of a mode of fixed $\ell$ 
peaks when the advection time is comparable to
the period of a lateral sound wave within the settling zone between the shock and the surface
of the accretor (see \S\ref{sec:mech}).  
The peak growth rate therefore moves to higher $r_*/r_\mathrm{s0}$ for
fixed $\ell$, or to a lower Legendre index for fixed $r_*/r_\mathrm{s0}$, 
as $\varepsilon$ is increased. This can be observed in Figs.~\ref{f:eigenfreq_e0.0} 
and \ref{f:epsilon_fundamental}a,c. 

The advective-acoustic cycle yields a WKB growth rate scaling as
$\omega_\mathrm{grow} \sim \ln{|Q|}/t_\mathrm{Q}$, where $Q$ is the 
efficiency of the cycle and $t_\mathrm{Q} \sim 2\pi/\omega_{\rm osc}$ its duration \citep{F07}. 
For larger dissociation energies, $t_\mathrm{Q}$ is dominated by the advection time, and
$t_\mathrm{Q}$ increases somewhat faster than linearly\footnote{For large values of $\varepsilon$, the density
profile steepens relative to $\rho \sim r^{-3}$ and thus the velocity decreases faster than linearly with radius.
The advection time 
(\ref{eq:tadv}) is then dominated by the innermost part of the flow.} in $\kappa$.
The oscillation frequencies depend mainly on the flow time, as can be seen by scaling them to 
$|v_2|/(\rs0 - r_*)$ (Fig.~\ref{f:epsilon_fundamental}d).  On the other hand,
Fig.~\ref{f:epsilon_fundamental}c shows that the growth rates decrease with increasing $\varepsilon$
even after this first-order effect is removed.

The growth rate also depends on the
coefficients for the conversion of an outgoing sound wave to an ingoing entropy
or vortex wave at the shock; and for the linear excitation of an outgoing sound wave
by the ingoing mode. 
As is shown in Appendix
\ref{sec:shockcoeff}, the first coefficient {\it increases} with increasing dissociation energy.
In the limit of strong shock compression, it is given by eq.~(\ref{eq:rhos})
\begin{equation}
\frac{\delta\rho_2^S}{\delta \rho_2^-} = -\frac{2(\gamma-1)(1-\mathcal{M}_2)}{\mathcal{M}_2(1+\gamma\mathcal{M}_2)}
\sim \gamma^{1/2}(\gamma-1)\kappa^{1/2}.
\end{equation}
We therefore conclude that the reduction in growth
rate signals a decreasing efficiency for the conversion of an ingoing entropy-vortex perturbation
into a sound wave.   

To understand why this happens, note that the radial flow time increases
relative to the lateral sound travel time as $\varepsilon$ is pushed
upward.  This makes it more difficult
for the pressure perturbation at diametrically opposite points on the shock to maintain
the correct relative phase.
As a result, the peak in the growth rate is forced to
larger values of $r_*/r_{\rm s0}$.  
We also observe that the velocity gradient at the base of the settling flow becomes stronger with increasing shock compression:
the gradient scale 
$v_r/(dv_r/dr)$ 
is reduced by a factor of $\sim 2$ as $\varepsilon$ is
raised from 0 to $0.25v_{\rm ff}^2$.  We suspect that this second effect has a weaker influence on
the growth rate, given our observation  
that that rapid growth depends on an approximate equality between the radial 
flow time and the 
lateral
sound travel time in between the cooling layer and the shock
(see \S \ref{sec:mech}).

The fundamental mode can be stabilized for sufficiently high $\varepsilon$, as is apparent
in Fig.~\ref{f:epsilon_fundamental}. 
Nonetheless an unstable mode can always
be found among the higher radial overtones.  
Figure~\ref{f:epsilon_harmonic} shows the growth rates of the first 11 overtones of the $\ell = 1$ mode
as a function of $\varepsilon$, for $r_*/r_\mathrm{s0}=0.4$.  

Our ability to find linearly unstable modes stands in contrast to the analysis of \citet{yamasaki07}, which
employed a single, more realistic equation of state, and found no unstable modes in the absence of
neutrino heating.  
But the comparison between both studies is made difficult by the fact that we are keeping the
ratio $r_*/r_{\rm s0}$ fixed, whereas in the absence of heating \citet{yamasaki07} obtain a very
small shock radius. 
It should also be emphasized that our unstable modes undergo a weak non-linear development:
the position of the shock is  perturbed only slightly when 
$\varepsilon \ga 0.15\,v_{\rm ff}^2$ (see \S \ref{s:four}). Neutrino heating therefore plays a 
crucial role in maintaining large-amplitude oscillations of the shock
for realistic values of the dissociation energy $\varepsilon$ (Paper II).

A reduction in the upstream Mach number has a measurable effect on the frequency and growth rate
of the SASI (see Fig.~\ref{f:mach_eigenfreq}).  When $\mathcal{M}_1 > 10$, one reaches the
asymptotic, strong-shock regime, and the eigenfrequencies are essentially constant. 
Lower values of $\mathcal{M}_1$ push up the post-shock velocity and the real frequency of the mode (at fixed value
of $r_*/r_{\rm s0}$).
Growth is substantially reduced for a weak shock when $\varepsilon$ is close to zero,
but is only modestly affected when $\varepsilon$ is large.

The dependence of the growth rate ($\ell = 1$) on the adiabatic index $\gamma$ is shown in 
Fig.~\ref{f:gamma_harmonic} for $\varepsilon=0$ and $0.25v_{\rm ff}^2$,
in units of $\tff^{-1}$ as well as $|v_2|/(\rs0 - r_*)$.
Reducing $\gamma$ has some similar effects to increasing $\varepsilon$ at fixed $\gamma$:
the flow slows down below the shock, and the density profile steepens. 
The growth rate is therefore reduced,
and the peak growth rate is found at higher radial overtones (see \S{\ref{sec:mech}).  When the
radius of the shock is allowed to vary, the peak growth rate moves to a higher value of $r_*/r_{s0}$
as $\gamma$ is decreased.

\section{Nonlinear Development in Axisymmetry}
\label{s:four}

The non-linear development of the SASI 
is addressed in this section.
Our hydrodynamical simulations enforce axisymmetry, and 
ignore
the effects 
of neutrino heating.  Our main interest is in how nuclear dissociation limits the growth of
the SASI.  We first discuss the transition from a laminar accretion flow to a turbulent,
quasi-steady state.  We then explore some time-averaged properties of this state, and the 
global energetics of the flow below the shock.

\subsection{From Linear Instability to Saturation}

The oscillation of the shock, and the fluid below it, remains coherent during the 
initial linear phase.  Consider the $\ell = 1$ SASI mode.  The shocked fluid in the
expanding pole has an increased energy density relative to the surrounding fluid.
As a result, the expanding pole is the source of a sound wave that propagates laterally to the
other pole.  When the radial advection time is comparable to this lateral sound travel time,
this pressure enhancement reaches the opposite pole just as the phase of the oscillation
has reversed, thereby reinforcing the outward expansion of the shock.  The standing wave
observed by \citet{BM06} can be interpreted in this way (instead of the purely acoustic
phenomenon proposed by these authors).  Another visualization of the effect (for $\varepsilon=0$ 
can be seen in Movie 1 in the online material.

The non-linear development of the SASI, in the absence of nuclear dissociation or
in the presence of neutrino heating, involves relatively large shock deformations involving
non-radial flows and the development of strong shock kinks (e.g, \citealt{BM06,scheck08}). 
On the other hand, when nuclear dissociation takes a significant fraction of the accretion kinetic 
energy and the SASI is not forced by convection, shock oscillations saturate at a much lower 
amplitude, with shock kinks not forming. Nevertheless, nonlinear coupling between different
modes still takes place and the shock thus acts as a generator of vorticity. The transition
to the nonlinear saturated state for three different values of $\varepsilon$ is shown
in Movie 2 in the online material.

In the longer term, the axisymmetric flow reaches a quasi-steady state, with a broad range of
oscillation frequencies. 
One could draw a parallel between this behavior and that of confined 2D turbulence, which reaches a
quasi-steady state in which the vorticity accumulates on the largest spatial
scales, decaying on the very
long viscous timescale \citep{davidson}.  
The nonlinear saturation properties of the SASI have been studied in 2D by \citet{ohnishi06} and in 
3D by \citet{iwakami08}. Both of these studies include a number of effects, among them neutrino
heating and a particular semi-realistic equation of state. 
We defer an examination of the effects of heating
to Paper II, and focus here on the effects
of nuclear dissociation on this quasi-steady state.

\subsection{Time-Averaged Properties}

Since none of our simulated flows explode, we are able to average the properties of
the fluid over a fairly long period of time ($T \sim 10^2-10^3$ dynamical times at the unperturbed
position of the shock).
In a turbulent system that displays a quasi-steady state behavior, a time 
average is equivalent to an ensemble average (e.g. \citealt{davidson}).
The mean and r.m.s. fluctuation of a quantity $f(r,\theta,t)$ are defined
in the usual way,
\begin{equation}
\langle f(r,\theta)\rangle \equiv \frac{1}{T}\int_0^T\, f(r,\theta,t)\,dt,
\end{equation} 
and
\begin{equation}
\Delta f(r,\theta) \equiv \left( \langle f^2\rangle - \langle f \rangle^2\right)^{1/2}.
\end{equation}

Fig.~\ref{f:timeaveraged_profile} shows the velocity field 
and sound speed, averaged over time and angle $\theta$, in accretion 
flows with $\varepsilon = 0$ and $0.15v_{\rm ff}^2$.
At any given moment one can define minimum and maximum shock radii, which
allows the construction of time averages $\langle r_{s,\rm min}\rangle$, 
$\langle r_{s,\rm max}\rangle$ and 
fluctuations
$\Delta r_{s,\rm min}$, $\Delta r_{s,\rm max}$.
One can distinguish four different zones in the flow, from the inside out: (1) the cooling layer, 
extending from the inner boundary to roughly  the point of maximum of $\langle c_s \rangle$; 
(2) the adiabatic envelope, which is bounded above approximately by 
$\langle r_{s,\rm min}\rangle - \Delta r_{s,\rm min}$;
(3) the shock oscillation zone, extending from this radius out to
$\langle r_{s,\rm max}\rangle + \Delta r_{s,\rm max}$; and (4) the supersonic accretion flow.
It is immediately evident that, as expected from the quasi-steady state behavior, accretion proceeds 
almost the same as in the unperturbed case, with $\langle v_r \rangle$ and $\langle v_\theta \rangle$ 
closely following the unperturbed velocity profile. The fluctuations $\Delta v_r$ and $\Delta v_\theta$
are comparable in magnitude and much larger than the mean flow, although they remain 
below $\langle c_s\rangle$.
As expected, the angle average of $\langle v_\theta \rangle$ vanishes.
Notice also that the sound speed in the mean flow
is everywhere lower than in the initial configuration, a point that we explore below.
\begin{figure}
\includegraphics*[width=\columnwidth]{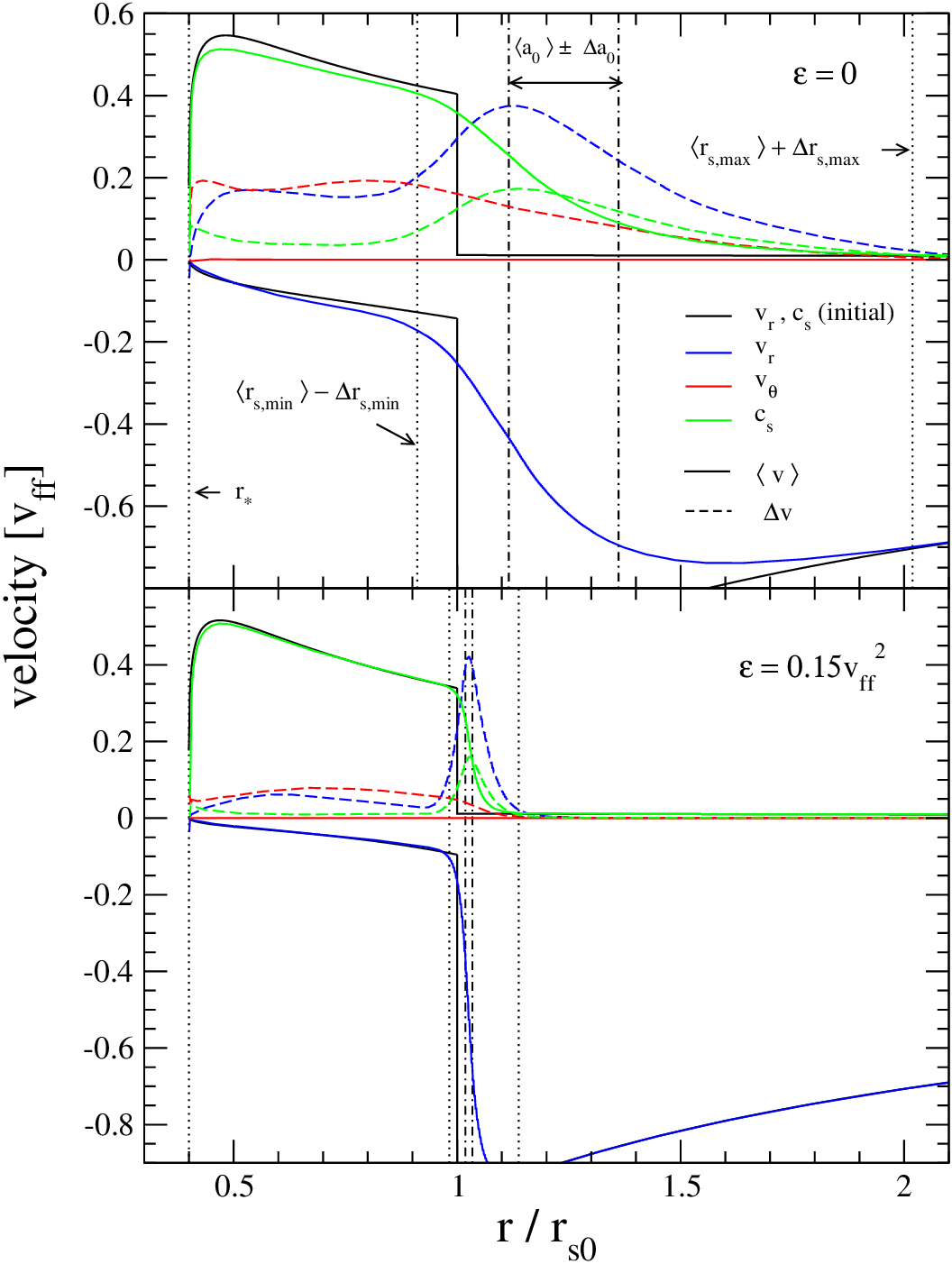}
\caption{Time-averaged profiles of velocity and sound speed in the fully developed, nonlinear phase 
of the SASI.  Upper panel represents a zero-energy flow ($\varepsilon=0$), and lower panel a flow
in which internal energy $\varepsilon=0.15v_\mathrm{ff}^2$ is removed below the shock.  Notice that the shock
moves over a smaller range of radii in the second case.  Spherical and
temporal averages are given by the solid lines, and r.m.s. fluctuations by the dashed lines, for:
radial velocity (blue), radial velocity in the unperturbed flow 
(lower black);
meridional velocity (red), sound speed (green)
and sound speed in the unperturbed flow 
(upper black)
Vertical lines denote: 
the inner boundary (left dotted); 
the inner excursion of the shock 
(middle dotted); the outer excursion of the shock (right dotted); the r.m.s. range of the 
$\ell = 0$ component of the shock radius 
(dot-dashed).  
See the text for definitions of these quantities.
Other parameters are $r_*/r_\mathrm{s0}=0.4$, $\gamma=4/3$, $\mathcal{M}_1 = 87$. 
}
\label{f:timeaveraged_profile}
\end{figure}

A dramatic feature of Fig.~\ref{f:timeaveraged_profile} is the reduction in the amplitude of
the shock oscillation as the effect of nuclear dissociation is introduced into the flow.  It is important
to understand the extent to which the oscillations seen in core
collapse simulations are the result of an intrinsic fluid instability, as opposed to a coupling
of the shock to the convective motions that are maintained by neutrino heating.
To this end, we have run a series of simulations with $\varepsilon$ increasing from 0 to 
$0.25\,v_{\rm ff}^2$, and analyzed the change in the character of the 
turbulence.  
Since the growth rate of the SASI is strongly
reduced by nuclear dissociation, we have chosen a ratio of accretion radius to shock radius
for which the growth rate 
of the fundamental
peaks when $\varepsilon = 0$ (namely $r_*/r_\mathrm{s0}=0.4$).
This is also compatible with our choice for studies of explosion hydrodynamics when neutrino
heating is included (Paper II).
A fraction $1 - 2\varepsilon/v_\mathrm{ff}^2$ of the accretion energy at the shock is available for
exciting turbulent motions below the shock.  This corresponds to $50\%-100\%$ of the accretion energy 
for the flows that we are examining.

The results of this study are summarized in Figs.~\ref{f:sat_radii_sample} and 
\ref{f:saturation_combined}, and Table~\ref{t:energies}.  The first figure
shows the amplitude of the $\ell=0$ and $\ell=1$ Legendre coefficients 
$a_{0,1}$ of the shock radius, as a function of time, for a few runs.  
Their amplitude drops dramatically
as $\varepsilon$ increases above $0.15\,v_{\rm ff}^2$.  
Some configurations display significant intermittency in the oscillations,
as exemplified with the case 
$\varepsilon = 0.1\,v_\mathrm{ff}^2$. 

The following figure shows the time-average $\langle a_{0,1}\rangle$
and r.m.s. $\Delta a_{0,1}$ of the Legendre coefficients 
($\langle a_1\rangle \approx 0$).
These quantities have been normalized to the time-average of the steady shock
position for a 1D 
run with the same parameters. This procedure allows us to remove a small
offset 
from $r_\mathrm{s0}$
in $\langle a_0\rangle$ that is caused by the finite initial velocity
at the inner boundary of the simulations,
as can be seen in the upper panels of Fig.~\ref{f:sat_radii_sample}.
The $\ell=0$ mode, being
stable, settles to a steady value within a few oscillation cycles.
For the lower values of $\varepsilon$, a modest monopole shift in the shock
radius remains in 2D, as compared with 1D, but this becomes insignificant
for $\varepsilon \geq 0.15 v_{\rm ff}^2$.  
The time-averaged $\ell = 1$ 
coefficient nearly vanishes when half of the accretion energy is
removed by dissociation ($\varepsilon = 0.25\,v_{\rm ff}^2$).

\begin{figure}
\includegraphics*[width=\columnwidth]{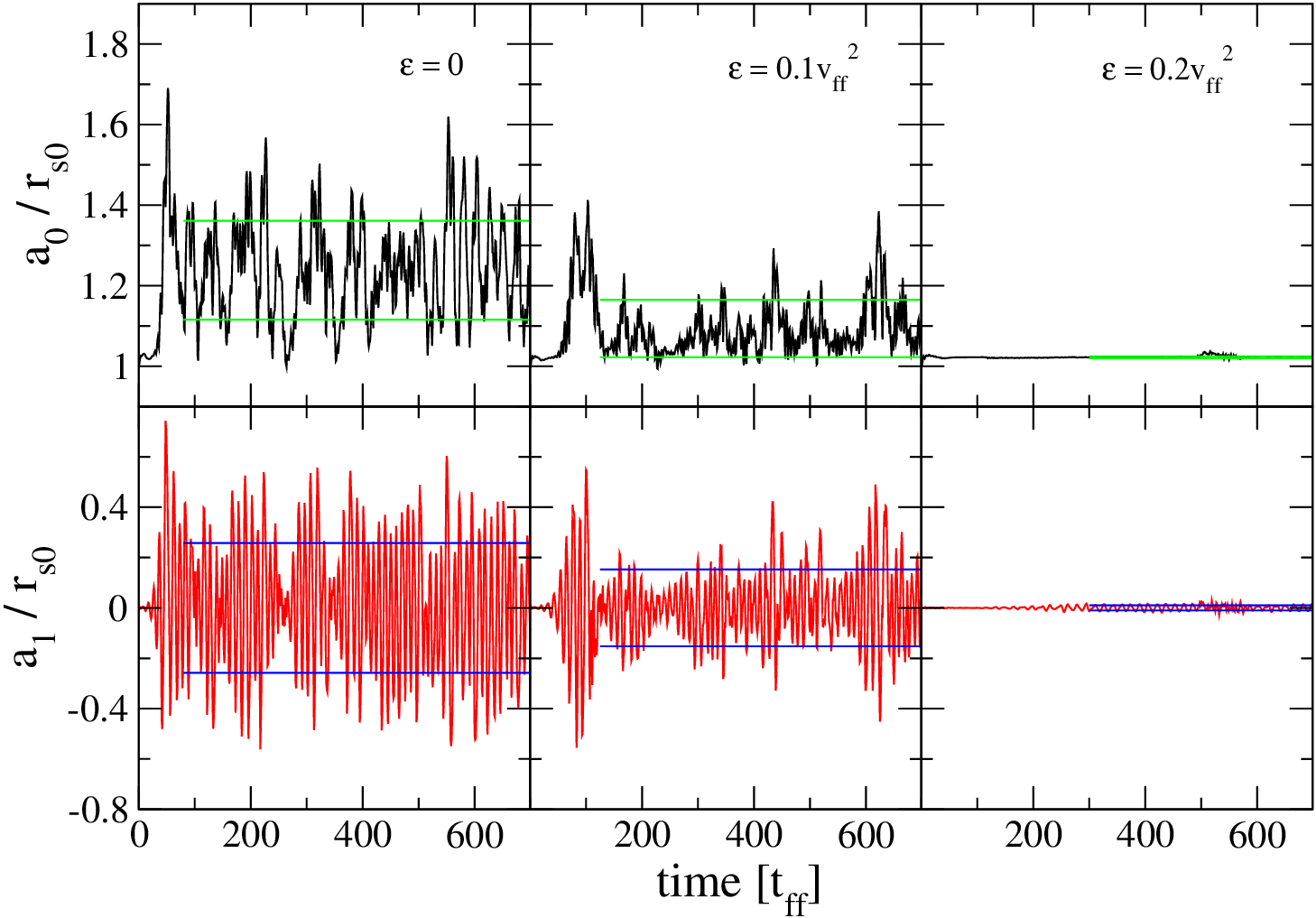}
\caption{Amplitudes of $\ell=0$ (upper) and $\ell=1$ (lower) components of the 
shock radius, versus time.  Panels from left to right represent flows with increasing
dissociation energies: $\varepsilon=0$, $0.1v_\mathrm{ff}^2$, and 
$0.2v_\mathrm{ff}^2$. The horizontal lines bound the r.m.s. 
fluctuation
of each Legendre coefficient.
These configurations correspond to those in Movie 2 (online material).
}
\label{f:sat_radii_sample}
\end{figure}

\begin{figure}
\includegraphics*[width=\columnwidth]{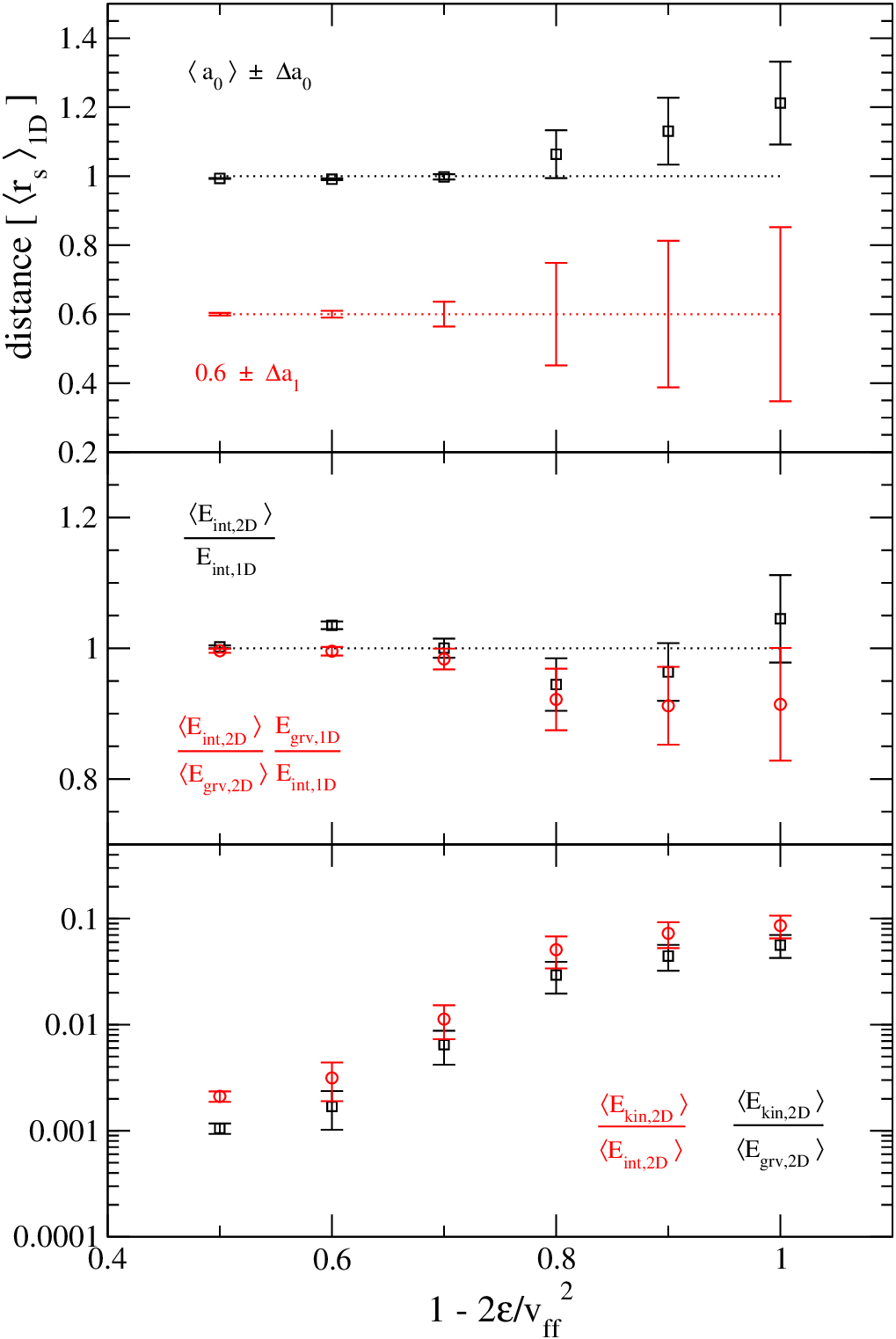}
\caption{Time-averaged properties of the shocked flow.  Horizontal axis labels the fraction
of the gravitational energy that is available for exciting oscillations of the shock.
Upper panel:  r.m.s. fluctuation
of the $\ell=0$ and $\ell=1$ components of the shock radius, 
with points denoting mean values.  Radius has been scaled to the equilibrium shock radius
in the 1D solution (which is stable).   Middle panel:  various components of the energy of
the 2D flow below the shock, averaged over time, relative to the 1D flow.
Black symbols denote internal energy and red symbols denote 
ratio of internal to gravitational energy.
Error bars represent the r.m.s. fluctuation.
Lower panel:  time average
of the fluid kinetic energy below the shock, relative to the internal energy (red
symbols) and gravitational energy (black symbols).  Notice the dramatic drop
in these ratios
with increasing dissociation energy.  
The system expands mainly as a result of the growth of turbulent kinetic energy, with a smaller
net change in internal energy.
}
\label{f:saturation_combined}
\end{figure}

In a real core collapse, the protoneutron star 
contracts by more than a factor of two in radius while the shock is
still stalled.  We have therefore run a series of simulations with
different $r_*/r_\mathrm{s0}$, to check whether the envelope size
has a significant influence on the properties of the saturated state.
Figure~\ref{f:saturation_combined_noepsilon} shows the 
$\ell=0,1$ Legendre coefficients for these runs, which have vanishing dissociation energy. 
As the envelope size increases by a factor of two, the fractional $\ell=0$ 
expansion of the shock nearly doubles, increasing from $1.15\langle r_s\rangle_\mathrm{1D}$ 
to $1.29\langle r_s\rangle_\mathrm{1D}$, 
where $\langle r_s \rangle_\mathrm{1D}$ is the steady shock position
in 1D. On the other hand, the rms amplitude of the $\ell=1$ oscillations 
only increases by $\sim 20\%$.

The partitioning of the energy of the fluid below the shock into different components is
explored in the lower panels of Figs.~\ref{f:saturation_combined} and \ref{f:saturation_combined_noepsilon}.
Since our 1D runs do not display any growing mode, and settle into a well-defined steady state configuration,
we use them to calculate reference values of the kinetic, internal, and gravitational potential energies.
These quantities are denoted by $E_{\rm kin,1D}$, $E_{\rm int,1D}$, and $E_{\rm grav,1D}$,
respectively,
with the spatial integral being carried out between  the radius at which $c_s$ peaks and $\langle r_s\rangle_{1D}$.
The analogous quantities $\langle E_{\rm kin,2D}\rangle$, $\langle E_{\rm int,2D}\rangle$,
and $\langle E_{\rm grav,2D}\rangle$ are obtained by 
integrating over volume and averaging over time.
\footnote{
The integration over volume is performed from the radius at which the angle and time 
average of $c_s$ peaks to the instantaneous shock surface.
}
Table~\ref{t:energies} displays the time averaged energies and their rms
fluctuations for both sequences of runs. By increasing $\varepsilon$ and thus the compression rate at the shock,
the mass in the adiabatic layer increases and so does the gravitational binding energy. The internal energy
also increases with increasing dissociation energy, because near hydrostatic equilibrium needs to be maintained. 
\begin{deluxetable}{cccccc}
\tablecaption{Time-Averaged Energy\tablenotemark{a} of Flow below Shock (2D)\label{t:energies}}
\tablewidth{0pt}
\tablehead{
\colhead{$r_*/r_\mathrm{s0}$} &
\colhead{$\varepsilon/v_\mathrm{ff}^2$} &
\colhead{$|\langle E_\mathrm{grv}\rangle|$} &
\colhead{$\langle E_\mathrm{int}\rangle$} &
\colhead{$\langle E_\mathrm{kin,r}\rangle$}&
\colhead{$\langle E_{\mathrm{kin},\theta}\rangle$}
}
\startdata
0.4   & 0    &  $62\pm 4$      & $40\pm 3$      & $1.8\pm 0.6$     & $1.7\pm 0.5$\\
      & 0.05 &  $74\pm 3$      & $45\pm 2$      & $1.6\pm 0.7$     & $1.7\pm 0.6$\\
      & 0.1  &  $95\pm 3$      & $55\pm 2$      & $1.4\pm 0.7$     & $1.4\pm 0.7$\\
      & 0.15 &  $129.1\pm 0.9$ & $74\pm 1$      & $0.4\pm 0.2$     & $0.4\pm 0.2$\\
      & 0.2  &  $200.0\pm 0.7$ & $107.6\pm 0.6$ & $0.2\pm 0.1$     & $0.1\pm 0.1$\\
      & 0.25 &  $320.3\pm 0.7$ & $159.6\pm 0.3$ & $0.21\pm 0.02$   & $0.13\pm 0.03$\\
0.5   & 0    &  $39\pm 3$     & $26\pm 2$     & $1.2\pm 0.5$     & $1.1\pm 0.4$\\
0.375 & 0    & $70\pm 6$      & $46\pm 4$     & $1.8\pm 0.7$         & $1.6\pm 0.5$\\
0.25  & 0    & $123\pm 7$    & $80\pm 4$     & $3.1\pm 1.0$         & $2.5\pm 0.7$
\enddata
\tablenotetext{a}{Energies are measured in units of $\rho_1 v_\mathrm{ff}^2 r_\mathrm{s0}^3 
\approx 3.4\times 10^{48} {\dot{M}}_{0.3} M_\mathrm{1.3}^{1/2} 
(r_\mathrm{s0}/150\mathrm{ km})^{1/2}$~erg.
Error bars denote the r.m.s. fluctuation.}
\end{deluxetable}

The internal energy remains almost constant, within fluctuations, relative to the 1D time averaged case, with some
increase for $\varepsilon=0$ and $\varepsilon=0.2$. Its ratio to the gravitational energy, however, 
is decreased relative to the 1D case by about 10\% for $\varepsilon=0$, about the
same fraction accounted for by the total kinetic energy. 
So our results point to a scenario in which the post-shock 
envelope expands as a result of the onset of turbulence, keeping its internal energy nearly constant. The 
fraction of the accretion energy that is converted to turbulent kinetic energy decreases with increasing $\varepsilon$,
from $\sim 10\%$ of the internal energy for $\varepsilon=0$ to about $0.1\%$ for $\varepsilon=2$. From Table~\ref{t:energies}
one can see that the total turbulent kinetic energy is very close to
the fluctuation in the internal energy. 
As regards the sequence for different $r_*/r_\mathrm{s0}$, we note that the fraction of the internal energy
going to turbulence decreases with increasing envelope size, but the absolute value of the turbulent kinetic energy
more than doubles.
\begin{figure}
\includegraphics*[width=\columnwidth]{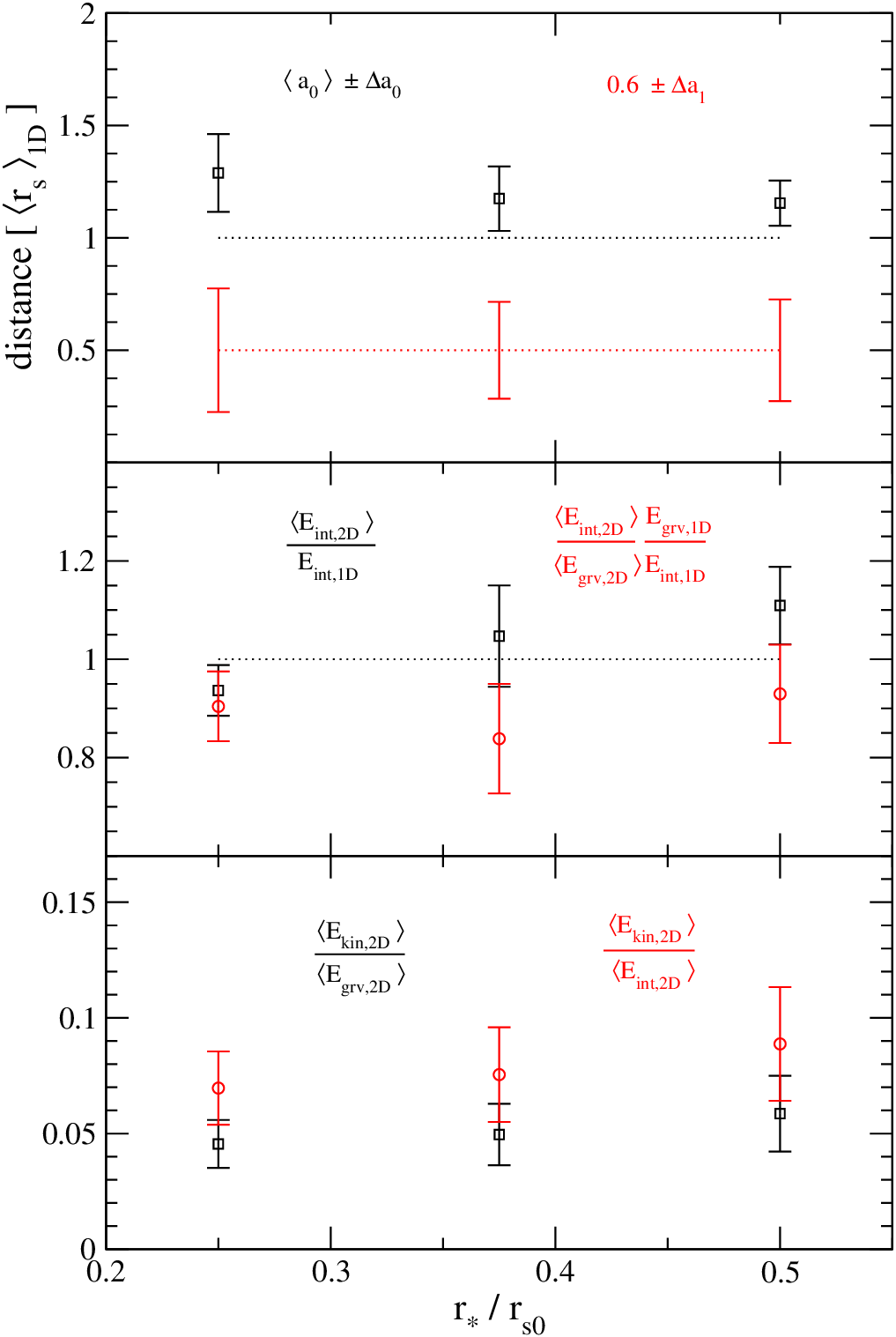}
\caption{Same as Fig.~\ref{f:saturation_combined}, but now varying $r_*/r_\mathrm{s0}$ for fixed $\varepsilon=0$.
Note that the vertical scale varies between the panels.}
\label{f:saturation_combined_noepsilon}
\end{figure}

\section{On the Linear Instability Mechanism}\label{s:five}\label{sec:mech}

The nature of the linear instability mechanism has been clearly demonstrated in the 
WKB regime \citep{F07}:  the growth of 
higher radial overtones is due to the ``advective-acoustic'' cycle first
described by 
\citet{foglizzo00}.  
This cycle involves a non-spherical deformation of the shock, which creates entropy and vortex waves 
in the downstream flow.
These perturbations become partly compressive as they are advected toward the star, 
thereby creating an 
outgoing  sound wave that interacts with the shock.  There is growth if the secondary shock oscillation
is larger in amplitude than the initial perturbation.   The instability is aided if the flow
is strongly decelerated somewhere below the shock.  There is no purely acoustic instability.

The linear stability analysis of 
\citet{F07} found approximate agreement with the real frequencies and growth
rates measured in the hydrodynamical simulation of \citet{BM06} -- which involve
low-order modes of the shock -- and good agreement for the $\ell=0$ modes in particular.  
\citet{yamasaki07} solved the eigenvalue problem in an accretion flow with a realistic equation
of state and strong neutrino heating.  They also found that the eigenfrequencies most
closely matched the advective-acoustic cycle.  
Hydrodynamical simulations
show that the oscillation period scales with the advection time at moderate to large
shock radius \citep{ohnishi06,scheck08}.

Here we provide further insight into the mechanism driving the fastest growing, 
low-frequency modes of a spherical shock.  We show that growth involves 
the radial advection
of entropy and vortex perturbations, and the lateral propagation of sound waves in the settling
flow below the shock.  This explains the observation by \citet{BM06} that the communication
of pressure perturbations by lateral sound waves plays a role in the SASI, but disagrees
with their inference that the mechanism may be purely acoustic.

To this end, we calculate the growth rate as a function of the size of the settling region
by solving the eigenvalue problem as formulated by \citet{F07}
for several spherical harmonics in a zero-energy accretion flow ($\varepsilon =0$).  
The basic configuration of the flow is the same as described in 
\S \ref{s:two}.  By changing
the ratio of $r_*$ to $r_{\rm s0}$, we are able to change the radial advection time relative
to the time for a sound wave to propagate laterally within the flow.  

The upper panel of Fig.~\ref{f:growth_timescales_fundamental} shows the growth rates of the 
fundamental (thick lines) and first radial overtone (thin lines) of the $\ell=1-4$ modes,  as a function of 
$r_*/r_\mathrm{s0}$.  The lower panel compares the advection time of the flow from the 
(unperturbed) shock in to $r_*$,
\begin{equation}\label{eq:tadv}
t_\mathrm{adv} = \int^{r_\mathrm{s0}}_{r_*} \frac{\ud r}{|v_r|},
\end{equation}
with the period of a meridional sound wave of Legendre index $\ell$.
The acoustic period is plotted at two different radii: right below the shock,
\begin{equation}
t_\mathrm{max,\ell} = \frac{2\pi r_\mathrm{s0}}{\ell c_{s,2}},
\end{equation}
and at the base of the settling flow,
\begin{equation}
t_\mathrm{min,\ell} = \frac{2\pi r_*}{c_\mathrm{s,*}} \approx t_\mathrm{max,\ell} \left( \frac{r_*}{r_\mathrm{s0}}\right)^{3/2}.
\end{equation}
Here $c_{s,2}$ is the sound speed downstream of the shock, and
$c_\mathrm{s,*} = c_{s,2}(r_\mathrm{s0}/r_*)^{1/2}$ is a good approximation to 
the sound speed at the top of the cooling layer.
The self-gravity of the accreting material is assumed negligible, and
most of the post-shock region is essentially adiabatic for the chosen cooling function. 
Hence, $t_\mathrm{max,\ell}$ corresponds to the longest possible period of a 
lateral sound wave of spherical harmonic $\ell$ outside the cooling layer.  The shortest acoustic period
($\sim t_\mathrm{min,\ell}$) is found at the base of the settling flow.
\begin{figure}
\includegraphics*[width=\columnwidth]{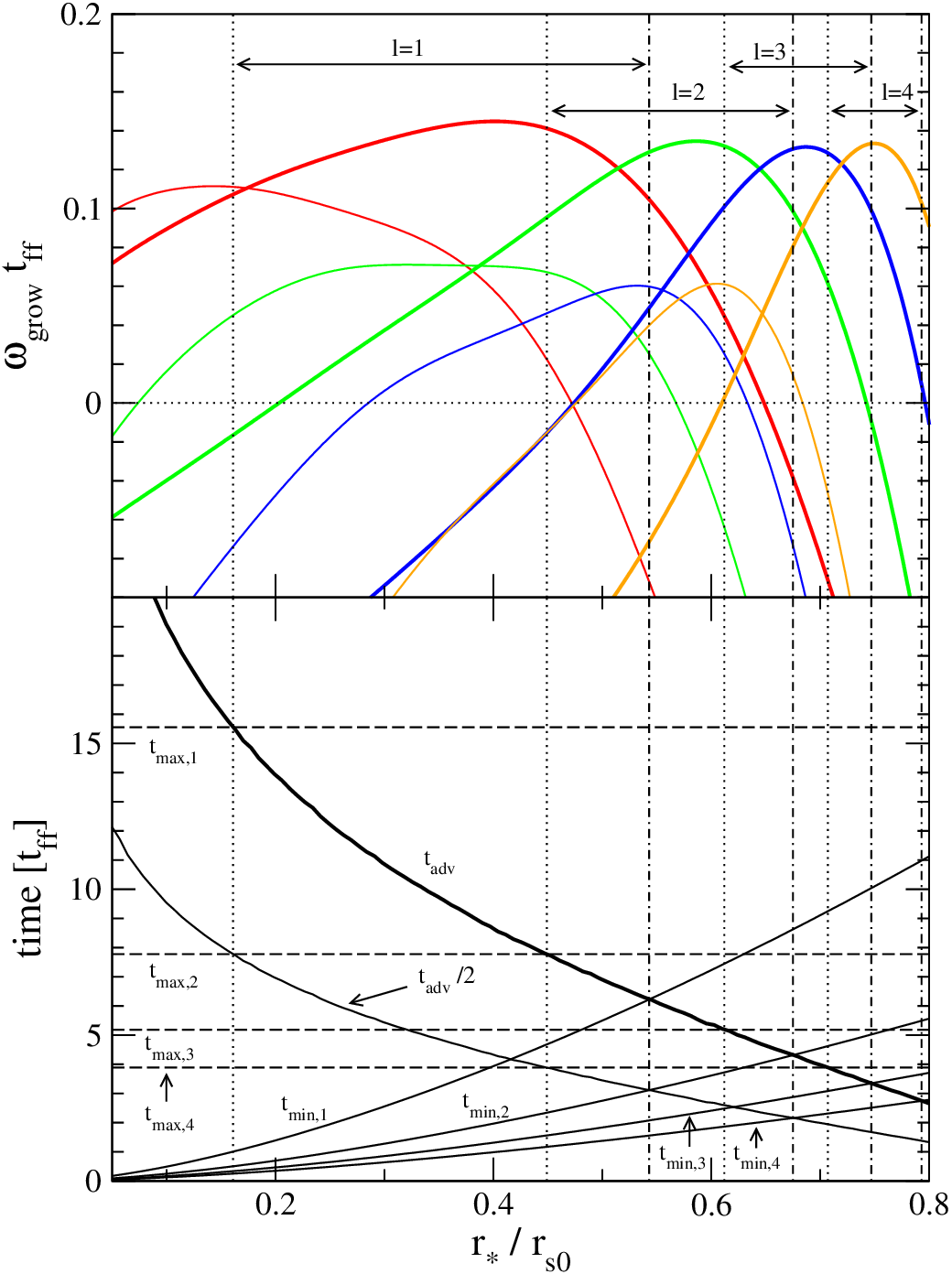}
\caption{Upper panel:  grow rates of linear modes of the shock, versus $r_*/r_\mathrm{s0}$ for
various spherical harmonics:  $\ell =1$ (red), $\ell=2$ (green), $\ell=3$ (blue) and $\ell=4$ (orange). 
Thick curves denote the fundamental mode; thin curves the first radial overtone.  
Lower panel: various timescales in the flow below the shock:  
advection time $t_\mathrm{adv}$ from shock to star; 
period $t_\mathrm{max,\ell}$ of lateral sound wave just below the shock; and
period $t_{\rm min,\ell}$ of lateral sound wave just above the cooling layer.
Dotted lines show the intersection between $t_\mathrm{adv}$ and $t_\mathrm{max,\ell}$, and dot-dashed lines
the intersection between $t_\mathrm{adv}$ and $t_\mathrm{min,\ell}$.  Growth of the fundamental
mode is concentrated where $t_\mathrm{min,\ell} < t_\mathrm{adv} < t_\mathrm{max,\ell}$, and
similarly $t_\mathrm{min,\ell} < t_\mathrm{adv}/(n+1) < t_\mathrm{max,\ell}$ for the $n^{\rm th}$ radial
overtone.
}
\label{f:growth_timescales_fundamental}
\end{figure}

For any given $\ell$, the peak of the growth rate of the 
fundamental mode is always found where $t_\mathrm{min,\ell} < t_\mathrm{adv} < t_\mathrm{max,\ell}$; the growth
rate of the first radial overtone peaks where $t_\mathrm{min,\ell}< t_\mathrm{adv}/2 < t_\mathrm{max,\ell}$.
The basic result does not depend
on whether one uses the radial advection timescale, or the sum of the advection time and the radial sound travel time
(which is significantly shorter).
Fig.~\ref{f:growth_timescales_harmonic} 
focuses on the radial overtones of $\ell=1$, which have significant growth at large values of
$r_\mathrm{s0}/r_*$. The peak of the n-th radial overtone satisfies 
$t_\mathrm{min,1} < t_\mathrm{adv}/(n+1) < t_\mathrm{max,1}$.
\begin{figure}
\includegraphics*[width=\columnwidth]{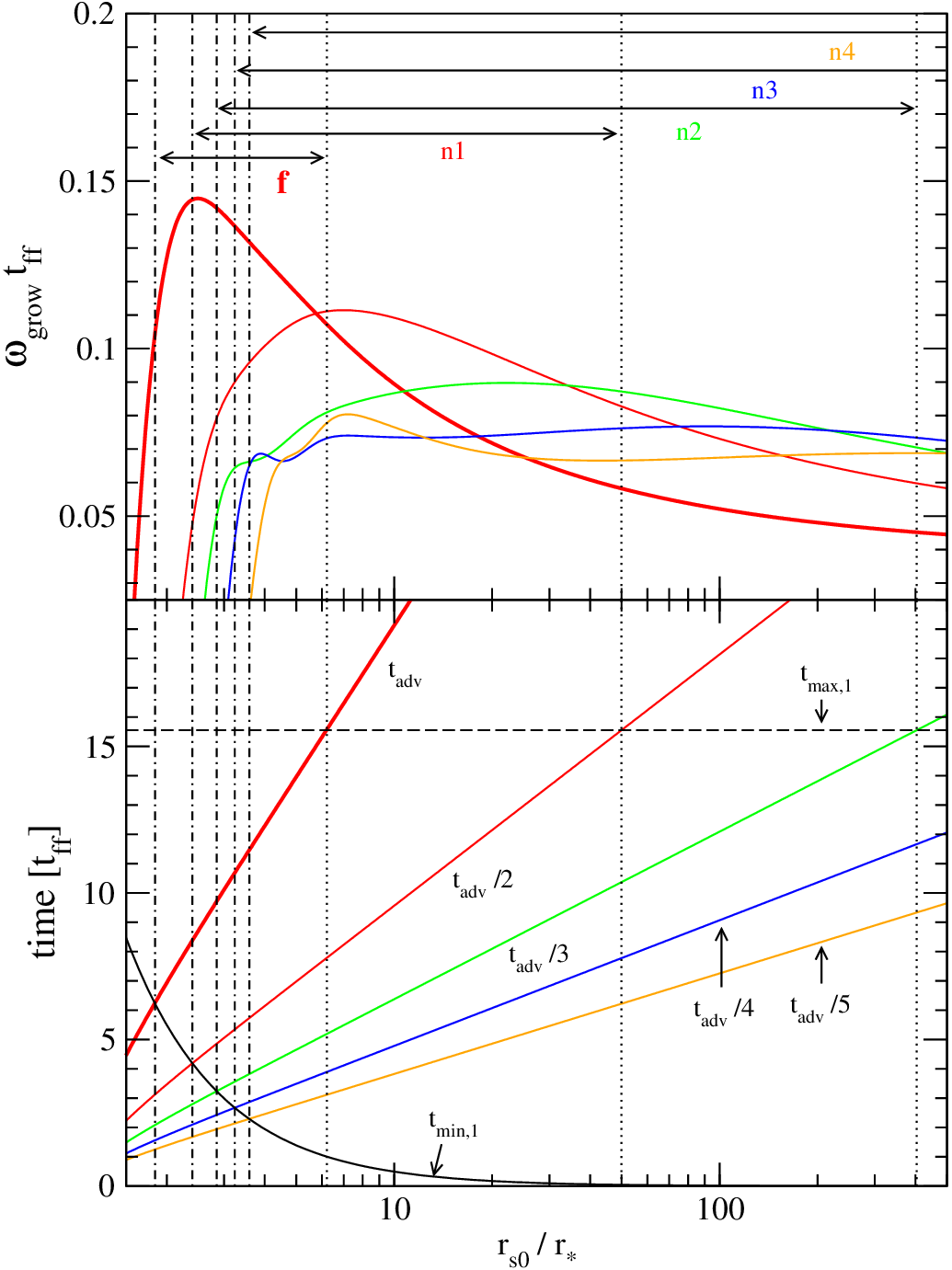}
\caption{Same as Fig.~\ref{f:growth_timescales_fundamental}, but now focusing on the radial overtones of $\ell=1$
at large $r_*/r_\mathrm{s0}$. We denote by {\bf f}, 
n1, n2, n3, and n4
the fundamental and first four overtones,
respectively. 
Other quantities have the same meaning as in Fig.~\ref{f:growth_timescales_fundamental}.
}
\label{f:growth_timescales_harmonic}
\end{figure}

The main conclusion that we draw from Figs.~\ref{f:growth_timescales_fundamental} and \ref{f:growth_timescales_harmonic}
is that the radial flow time controls the overtone of the fastest-growing mode, while the period of the meridional sound wave
controls the angular order.  
This interplay between the two timescales points to the advective-acoustic cycle as the mechanism
driving the instability.

The $\ell=0$ mode deserves special comment.  We find that its oscillation period is nearly twice the
duration of the advective acoustic cycle, $2(t_\mathrm{adv} + t_\mathrm{s,up})$, where $t_\mathrm{s,up}$ is the 
time taken for a sound wave to travel 
radially
from $r_*$ to $r_\mathrm{s0}$.   This is demonstrated for a wide
range of shock radii in Fig.~\ref{f:F07_L0_period}.  A purely spherical perturbation of the shock
will excite a downgoing entropy wave.  When the shock oscillation reaches its maximum or minimum radius, the sign of 
the pressure perturbation below the shock is generally opposite to that of the entropy perturbation,
as may be seen by combining eqs. [\ref{eq:delta_def}], [\ref{eq:sper}], [\ref{eq:drhos}], and [\ref{eq:rhos}].
When the entropy wave reaches the cooling later, 
the sign of the change in the (negative) cooling rate per unit volume (eq.~[\ref{eq:cool0}]) is
the same as the sign of the entropy perturbation
(at constant pressure),
$\delta \mathscr{L}_0/\mathscr{L}_0 = -[(\gamma-1)/\gamma](\beta-\alpha)\delta S$.
As a result, a positive pressure perturbation at
the shock\footnote{As measured at its unperturbed position.} generates an increase in the cooling rate
at the base of the settling flow, and therefore a negative pressure perturbation that is communicated
back to the shock on the radial acoustic time.  This secondary pressure
perturbation is in phase with the pressure perturbation at the shock if the mode period is twice the
period of the advective-acoustic cycle. This analysis of the $\ell=0$ mode helps to explain the importance 
of lateral acoustic waves in the growth of non-radial perturbations. 
\begin{figure}
\includegraphics*[width=\columnwidth]{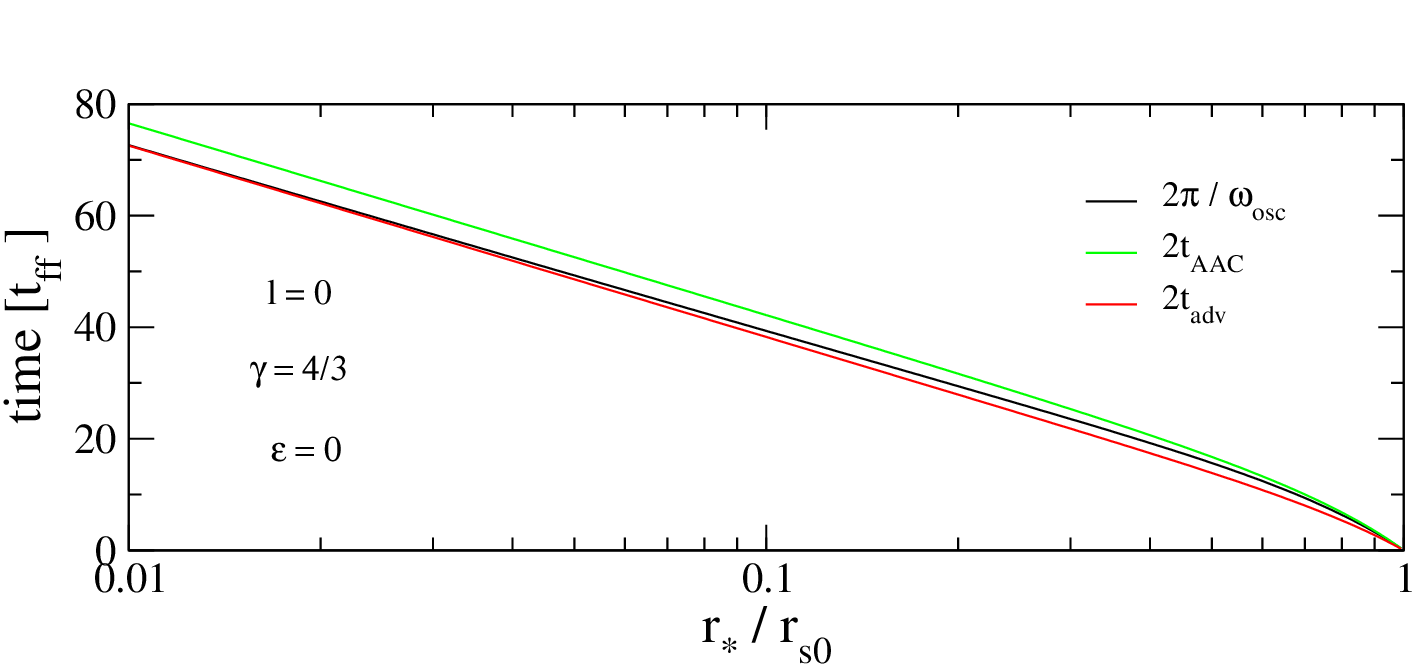}
\null
\caption{Period of the $\ell=0$ fundamental mode (black curve), which is very close
to twice the advection time (red) and twice the duration of the advective-acoustic cycle (green).
Parameters of the flow: $\gamma=4/3$ and $\varepsilon=0$.}
\label{f:F07_L0_period}
\end{figure}

\section{Summary}\label{s:six}

We have investigated the linear stability, and non-linear evolution, of 
a standing shock in a spherical accretion flow onto a compact star,
in the context of core-collapse supernovae.
Our analysis combines two-dimensional, time-dependent hydrodynamic simulations 
with a linear stability analysis.  The equation of state used in these calculations
allows for a loss of internal energy to nuclear dissociation below the shock, but
is otherwise that of an ideal gas.  We do not include heating effects associated
with charged-current absorption of neutrinos on protons and neutrons, or the
recombination of alpha particles.  A companion paper explores how the threshold
for an explosion is influenced by these additional effects (Paper II).

Our main results are as follows:

\noindent 1. --- The growth rate of the Standing Accretion Shock Instability (SASI) is significantly
reduced when $\sim 20-50$\% of the gravitational binding energy of the flow is
absorbed by nuclear dissociation.  A lowering of the adiabatic index of the flow
(or, equivalently, a steepening of the density profile above $\rho(r) \sim r^{-3}$)
has a similar effect.   
The lower growth rates appear to result from
a weaker coupling between the ingoing entropy-vortex
wave and a sound wave that acts back on the shock.
This reduced coupling is caused in part by 
a lengthening of the radial flow time in comparison with
the time for a sound wave to propagate laterally below the shock.
Increasing the dissociation energy also pushes the 
strongest instability to lower $\ell$ at fixed $r_*/r_{s0}$, and to higher $n$ at 
fixed $\ell$.

\noindent 2. --- By solving the eigenvalue problem, we do find some unstable modes for all
values of $\varepsilon$.  By contrast, the linear stability analysis of
\citet{yamasaki07} found no unstable modes for radial overtones $\leq 2$ at vanishing
neutrino luminosity (but evidence for very strong growth of $n = 3$ modes, which we
do not find),
albeit with small shock radii since their cooling function was fixed.

\noindent 3. --- 
The nonlinear development of the SASI in two dimensions is characterized by a
quasi-steady state that is reached after the linear modes have saturated.   As the dissociation
energy is increased, the fluctuations of the shock about its unperturbed position are greatly
reduced in amplitude.  We infer that strong neutrino heating is
needed to drive large-amplitude, dipolar oscillations of the shock in a realistic
core collapse environment.  

\noindent 4. --- The expansion of the shock that we observe
appears to be driven mainly (but not entirely) by the action of turbulent pressure; 
we observe a smaller absolute positive change in the internal energy of the flow
below the shock. 
The turbulent kinetic energy saturates at $\lesssim 10\%$ of the 
internal energy.  A large expansion of the shock therefore is possible only if
the flow has nearly vanishing total energy.

\noindent 5. --- The strongest growth of shock perturbations of spherical harmonic $\ell$ 
and radial overtone $n$ is encountered when the radial flow time across a distance
$\sim (r-r_*)/(n+1)$ equals the period 
$2\pi r/\ell c_s(r)$ 
of a lateral sound wave, at some radius
$r$ in between the cooling layer and the shock.  This provides a 
compelling argument as to why the linear instability involves a feedback between a 
radially propagating entropy-vortex perturbation, and a laterally propagating
acoustic perturbation.  

\noindent 6. --- The period of a spherical ($\ell = 0$) mode
is nearly twice the radial flow time,
that is, approximately double the period of the fundamental non-radial modes.
This is due to a change in sign between the pressure perturbation at the shock, and
the pressure perturbation that is induced in the cooling layer at the base of
the accretion flow
by the advected entropy perturbation.


\acknowledgments
We are grateful to Jonathan Dursi
for scientific discussions and help with FLASH.
CT acknowledges some stimulating discussions of 
the shock stability problem with Ethan Vishniac. 
Careful and constructive comments by an anonymous referee helped improve this article.
The software used in this work was in part developed by the DOE-supported ASC / Alliance 
Center for Astrophysical Thermonuclear Flashes at the University of Chicago.
Computations were performed in the CITA
Sunnyvale cluster, which was funded by the Canada Foundation for Innovation. 
This research was supported by NSERC of Canada. 
RF is supported in part by the Ontario Graduate Scholarship Program.

\appendix

\section{A. Numerical Treatment of Nuclear Dissociation in FLASH2.5}
\label{s:burning}

Nuclear dissociation is implemented by means of the \emph{fuel+ash} nuclear burning module in FLASH2.5 \citep{fryxell00}.
To prevent undesired dissociation effects upstream of the shock, we replace the density and temperature
thresholds for burning in 
the default version of the code
with a threshold in Mach number:  burning takes place so long as the
fluid has a Mach number lower than $\mathcal{M}_\mathrm{burn} = 2$.   This has the effect of localizing
the nuclear dissociation right behind the shock.
We find a very limited amount of incomplete burning whenever the shock is strongly deformed,
so that it runs nearly parallel to the upstream flow.   Of course, a similar phenomenon is expected
in the full collapse problem (incomplete dissociation behind a strongly oblique shock leading
to
the formation of cold and fast accretion plumes);  so we have made no attempt to completely eliminate this effect.

In principle all of the
kinetic energy of accretion can be absorbed by nuclear dissociation downstream of a shock:
the Riemann problem has a limiting solution $\rho_2\to \infty$ and $v_2\to 0$, with
$\rho_2 v_2 = \rho_1 v_1$, which for a strong shock yields
$\varepsilon \to v_1^2/2$.
However, the numerical evolution of the problem in discrete timesteps limits the dissociation
energy to be smaller in magnitude than the specific internal energy of the fluid, otherwise
negative pressure results. In particular, since we want to maintain a steady shock as
background flow for stability calculations, the maximum dissociation energy that can be
removed in a single timestep at $r=r_\mathrm{s0}$ is 
the internal energy
of the postshock flow
after dissociation has been subtracted.
That is, if we allow for the density contrast $\kappa$ to increase according to eq.~(\ref{eq:kappa_phot}),  then we
require
\begin{equation}
\label{eq:epsilon_maxnum}
\varepsilon < \frac{v_1^2}{\gamma(\gamma-1)\mathcal{M}_2^2\kappa^2},
\end{equation}
where $\mathcal{M}_2$ is the post-shock Mach number, eq.~(\ref{eq:m2}).
The maximum single-step dissociation energy $\varepsilon_\mathrm{max}$ is obtained by equating the two sides
of eq. (\ref{eq:epsilon_maxnum}).  One finds
$\varepsilon_\mathrm{max} \simeq 0.213v_\mathrm{ff}^2$ for
$\gamma = 4/3$ and
$\mathcal{M}_1 \to \infty$.

In reality, burning of the shocked fluid occurs within a layer of a finite width.
To relax the above limit on $\varepsilon$, we have modified the default \emph{fuel+ash} submodule
of FLASH2.5 to allow nuclear
dissociation to occur in stages:  instead of burning all the \emph{fuel} into \emph{ash} in a single step,
we only allow the burning of a fraction
$1/n_\mathrm{burn}$ at a time. The value of $n_\mathrm{burn}$ is adjusted empirically so as to avoid
numerical problems at large expansions of the shock, where
$\varepsilon$ may approach the local value\footnote{As we discuss in \S~4, the
shock oscillations saturate at a low amplitude when 
$\varepsilon > 0.15v_\mathrm{ff}^2$, 
which means that this
effect is negligible for the simulations reported in this paper.}
of $v_{\rm ff}^2/2$.
In most cases, the burning is spread across $\sim n_\mathrm{burn}$ cells behind the shock.  We have
checked that the mode frequencies of the flow are insensitive to the particular choice,
as long as the single-step constraint is met.

\section{B. Procedure for Obtaining Eigenfrequencies from a Hydrodynamical Simulation}
\label{sec:eigenmeasure}

Here we describe the procedure to obtain linear eigenfrequencies from our time
dependent numerical simulations, and the associated error.  
To excite a particular mode in our simulations, we add an overdense shell upstream of the shock
with the desired angular dependence and 
a sinusoidal radial profile (amplitude $\sim 10$~\%, width $\sim 0.2r_\mathrm{s0}$). 
This generates an 
initial shock displacement with very small amplitude (negligible compared with the spherically symmetric
displacement described in \S~\ref{sec:time_dep}), followed by a growth of the SASI mode on the advection
timescale. 

To measure the growth rate $\omega_\mathrm{grow}$ and oscillation frequency $\omega_\mathrm{osc}$ 
of a mode associated with a given Legendre-$\ell$, we project the shock surface onto the corresponding 
Legendre polynomial, obtaining a time-dependent Legendre coefficient,
\begin{equation}
\label{eq:a_l}
a_\ell(t) = \frac{2\ell+1}{2}\int_0^\pi\, R_\mathrm{s}(\theta,t)\,P_\ell(\theta) \sin{\theta}\,d\theta,
\end{equation} 
where $P_\ell$ is the Legendre polynomial and $R_\mathrm{s}(\theta,t)$ is the shock surface, defined as
the locus of points with pressure $p_\mathrm{shock} = \sqrt{p_1 p_2}$. For the strong 
shock simulations, this has a numerical value $p_\mathrm{shock} \simeq 0.01\rho_1 v_\mathrm{ff}^2$.

The time-dependent Legendre coefficient is then fitted with the functional form
\begin{equation}
a_\ell(t) = c_1 + c_2\, e^{\omega_\mathrm{grow}t}\,\sin{(\omega_\mathrm{osc}t + c_3)} + c_4t,
\end{equation}
where $\omega_\mathrm{grow}$ is the growth rate, $\omega_\mathrm{osc}$ the oscillation frequency, 
and the $c_i$ are constants. This fitting function works well whenever there is only a single unstable 
harmonic, usually the fundamental, for a given $\ell$ (see, e.g., lower-left panel in Fig.~\ref{f:sat_radii_sample} during
the first $\sim 100t_\mathrm{ff}$). 
The fitting is performed with a Levenberg-Marquardt algorithm \citep{NR}.

For the fitting we choose a temporal range that begins when the Legendre coefficient starts a clean sinusoidal 
oscillation, and which ends when $a_\ell(t) = (r_\mathrm{s0}-r_*)/10$ for $\ell \geq 1$, corresponding to a shock displacement 
of $10\%$ of the unperturbed value at the poles. Our temporal sampling interval is $t_\mathrm{ff}/2$, about a quarter of 
the radial sound crossing time 
from $r_*$ to $r_\mathrm{s0}$.

To estimate the uncertainty in the fitted eigenfrequencies, we assign a ``measurement error'' to 
$a_\ell(t)$, which we compute as follows. The error in $R_\mathrm{s}(\theta,t)$ is taken to be 
one-half the size of the baseline resolution, $\Delta r_\mathrm{base}/2$.
The size of the angular cell 
$\Delta \theta_\mathrm{base}$
enters through the computation of the integral in eq.~(\ref{eq:a_l}) as a discrete sum. Errors adding up in
quadrature then yield
\begin{equation}
\label{eq:}
\delta a_\ell = \frac{2\ell + 1}{4}\Delta r_\mathrm{base} \Delta \theta_\mathrm{base} \sqrt{\sum_i P_\ell^2(\theta_i)\,\sin^2{\theta_i}},
\end{equation}
where the sum is performed over all the angular cells. This result is then used as the input error in the 
Levenberg-Marquardt algorithm. The error bars shown in Fig.~\ref{f:convergence_complete} are the 1-sigma 
errors that output from the fitting routine, multiplied by 3.

\section{C. Linear Stability Analysis with Nuclear Dissociation Downstream of the Shock}
\label{sec:bndcnd_epsilon}

We outline our approach to calculating linear perturbations of
a spherical accretion flow with a standing shock wave.
The surface of the star imposes a hard inner boundary to the flow, and therefore introduces a
free parameter $r_*/r_{\rm s0}$, the ratio of the stellar radius to unperturbed shock radius.
The structure of the flow is further defined by the equation of state and the cooling function.  We generalize
the ideal gas equation of state to include a finite, uniform dissociation energy $\varepsilon$
downstream of the shock, as is described in \S \ref{sec:initial_cond}.
We explicitly account for the low-entropy cutoff in the cooling function (eq.~[\ref{eq:cooling_cutoff}]).

\citet{F07} formulate this eigenvalue problem assuming a constant-$\gamma$ ideal gas equation
of state.  
Generalizing this formalism to account for a constant dissociation energy $\varepsilon$ is
straightforward when thermal energy is removed from the flow only immediately below the shock.
The background flow solution is modified, but the algebraic form of the perturbation equations
is not.  We therefore use the differential system given in eqns.~(10)-(13) of \citet{F07}, 
which employs the variables
$f = v_r \delta v_r + 2c_s \delta c_s/(\gamma-1)$, $h = \delta(\rho v_r)/\rho v_r$,
$\delta S = (\delta p/p - \gamma \delta \rho/\rho)/(\gamma-1)$,
and
$\delta K = r^2 {\bf v}\cdot(\bnabla\times\delta\bomega) + \ell(\ell+1)(c^2/\gamma)\delta S$,
where $\bomega = \bnabla \times {\bf v}$.
All quantities are projected onto spherical harmonics
$Y_\ell^m$. 
A complex eigenvalue
$\omega \equiv \omega_\mathrm{osc} + i\omega_\mathrm{grow}$ 
is obtained by enforcing $\delta v_r = 0$ at $r=r_* + 10^{-4}(r_\mathrm{s0}-r_*)$.
The eigenfrequencies obtained are nearly insensitive to the radius at which this boundary conditions is applied, so long as
it lies inside the cooling layer. 

A subtlety in the treatment of the boundary conditions at the shock is worth discussing.
When the shock is perturbed to a position $r_{s0} + \Delta r$ and velocity $\Delta v$,
the equation of energy conservation across the discontinuity becomes
\begin{equation}\label{eq:en}
{1\over 2}(v_1 - \Delta v)^2 + {c_{s1}^2\over\gamma -1} =
{1\over 2}\left(v_2 + \delta v_r - \Delta v\right)^2 + 
{\gamma\over\gamma-1}\left(\frac{p_2}{\rho_2} + {\delta p_2\over\rho_2} - {p_2\over\rho_2}{\delta\rho_2\over\rho_2}\right)
+ {\partial \varepsilon\over\partial\rho}\delta\rho_2 + {\partial\varepsilon\over\partial p}\delta p_2.
\end{equation}
Here, as before, $1$ and $2$ label the upstream and downstream flows in the frame of the accretor.
When $\varepsilon = $ constant, the algebraic form of this equation is unchanged from the case
$\varepsilon = 0$.  Although the Bernoulli parameter $b$ of the background flow below the shock
is reduced, the perturbation $f = \delta b$ does not
receive additional terms.  The boundary conditions on $f$, $h$, $\delta S$, and $\delta K$ therefore have
the same algebraic form as eqns.~(B10)-(B12), (A6) and (B15) of \citet{F07}. By contrast,
the shock boundary conditions on the perturbation variables
$\delta\rho_2$, $\delta p_2$ and $\delta v_r$ do acquire additional terms resulting from the 
changing compression ratio $\kappa$ (eq. [\ref{eq:kappa_phot}]).
We have checked that these additional terms cancel
out in the boundary conditions on $f$, $h$, $\delta S$, and $\delta K$.

To account for our entropy cutoff (eq.~\ref{eq:cooling_cutoff}) in the linear stability calculation, one needs to add
an additional term to the perturbation to the cooling term in the 
energy equation,
\begin{equation}
\delta \left( \frac{\mathscr{L}}{\rho v}\right) = \left(\frac{\mathscr{L}}{\rho v}\right)
\left[ (\beta-1)\frac{\delta \rho}{\rho} + \alpha\frac{\delta c^2}{c^2} -\frac{\delta v_r}{v_r}
-2\frac{s}{s_\mathrm{min}^2} \right].
\end{equation}
The precise analytic eigenfrequencies are somewhat sensitive $s_\mathrm{min}$, so this parameter needs to be the same
in both linear stability and simulation for proper comparison.
The difference between including and excluding the entropy cutoff in the linear stability
can be seen from Fig.~\ref{f:eigenfreq_e0.0}, where results obtained without entropy cutoff are shown as dashed lines.

\section{D. Coefficient Relating the Amplitude of an Ingoing Entropy
Wave to an Outgoing Sound Wave at a Standing Shock:  Effects of
Nuclear Dissociation}
\label{sec:shockcoeff}

We focus here on the WKB limit, where the wavelength of the excited modes is small compared with the density scale
height below the shock.  The perturbed equations of mass, momentum and energy
conservation 
across the shock
are
\begin{equation}\label{eq:mass}
\rho_1 (v_1 - \Delta v) = (\rho_2 + \delta\rho_2) 
(v_2 + \delta v_r - \Delta v),
\end{equation}
\begin{equation}\label{eq:mom}
\rho_1 (v_1 -\Delta v)^2 + p_1
= (\rho_2 + \delta\rho_2)(v_2 + \delta v_r - \Delta v)^2 + p_2 + \delta p_2,
\end{equation}
and eq. (\ref{eq:en}) with $\partial\varepsilon/\partial\rho = 
\partial\varepsilon/\partial P = 0$.
Since the end result becomes fairly complicated in the case
of non-vanishing dissociation energy, we now simplify the discussion by
taking the limit ${\cal M}_1 \rightarrow \infty$.  
Working to linear order in the perturbation variables, we divide up them up into contributions
from an outgoing sound wave $-$ (moving oppositely to the background flow), an ingoing sound wave
$+$, and an entropy perturbation $S$.  Then
\begin{equation}
\label{eq:delta_def}
\delta p_2^\pm = c_{s2}^2 \delta\rho_2^\pm;
\qquad
\delta v_r^\pm = \mp c_{s2} {\delta\rho_2^\pm\over \rho_2};
\end{equation}
\begin{equation}\label{eq:sper}
\delta S_2 = -{\gamma\over\gamma-1}{\delta\rho_2^S\over\rho_2};\qquad
\delta p_2^S = \delta v_r^S = 0.
\end{equation}
Equation (\ref{eq:mom}) becomes 
\begin{equation}
\label{eq:drhos}
\delta\rho_2^+ = - \frac{(c_{s2}+v_2)^2}{(c_{s2}-v_2)^2}\delta\rho_2^- 
                 - \frac{v_2^2}{(c_{s2}-v_2)^2}\delta\rho_2^S
               = - \frac{(1-\mathcal{M}_2)^2}{(1+\mathcal{M}_2)^2}\delta\rho_2^- 
                 - \frac{\mathcal{M}_2^2}{(1+\mathcal{M}_2)^2}\delta\rho_2^S
\end{equation}
where ${\cal M}_2 = |v_2|/c_{s2} = 1/\sqrt{\gamma(\kappa-1)}$.
Substituting for $\delta\rho_2^+$, equations (\ref{eq:en}) and (\ref{eq:mass}) become 
\begin{equation}
-(\rho_2-\rho_1)v_1\Delta v = 2\frac{(1-\mathcal{M}_2)}{(1+\mathcal{M}_2)}\mathcal{M}_2 c_{s2}^2 \delta \rho_2^-
                          - \frac{\left[1+\mathcal{M}_2+(\gamma-1)\mathcal{M}_2^2\right]}{(1+\mathcal{M}_2)}
                            \frac{c_{s2}^2}{(\gamma-1)}\delta\rho_2^S
\end{equation}
and
\begin{equation}
(\rho_2 - \rho_1)\Delta v = 2\frac{(1-\mathcal{M}_2)}{(1+\mathcal{M}_2)}c_{s2}\delta\rho_2^-
                            -\frac{\mathcal{M}_2}{(1+\mathcal{M}_2)}c_{s2}\delta\rho_2^S
\end{equation}
respectively. One also has 
\begin{equation}
v_1 = -c_{s2}\left({\cal M}_2 + {1\over \gamma{\cal M}_2}\right),
\end{equation}
for arbitrary $\kappa > 1$.  Combining these equations gives
\begin{equation}\label{eq:rhos}
\frac{\delta\rho_2^S}{\delta\rho_2^-} = -\frac{2(\gamma-1)(1-\mathcal{M}_2)}{\mathcal{M}_2(1+\gamma\mathcal{M}_2)}
\end{equation}
This result can be expressed in terms of the perturbed
Bernoulli variable
[$f^S = - (\gamma-1)^{-1}\,c_{s2}^2\,\delta\rho_2^S/\rho_2$;
$f^\pm = (1\pm {\cal M}_2)c_{s2}^2\,\delta\rho_2^\pm/\rho_2$],
\begin{equation}
{f^S\over f^\pm} = {2\over{\cal M}_2(1+\gamma{\cal M}_2)}.
\end{equation}
When the dissociation energy vanishes,
${\cal M}_2 = [(\gamma-1)/2\gamma]^{1/2}$ and eq. (\ref{eq:rhos}) reduces to 
\begin{equation}
{\delta\rho_2^S\over\delta\rho_2^-} = -{4{\cal M}_2\over 1+ 2{\cal M}_2}.
\end{equation}
In this particular case, we find agreement with the coefficient
derived in \citet{foglizzo05} (their equation [F11]).
That previous result cannot, however, be used when $\varepsilon \neq 0$,
because its derivation depends on the Hugoniot adiabatic \citep{landau}.
One can also show that, in the absence of dissociation,
$\delta \rho_2 = \delta\rho_2^+ + \delta\rho_2^- + \delta\rho_2^S = 0$.
This is no longer the case when $\varepsilon\neq 0$, because a net density perturbation
arises due to the change in the compression rate [eq.~(\ref{eq:kappa_phot})], $\delta\rho_2 = \rho_1 \delta\kappa$, 
with
\begin{equation}
\delta\kappa = -\frac{\partial \kappa}{\partial v_1}\Delta v.
\end{equation}

\bibliographystyle{apj}
\bibliography{references,apj-jour}

\end{document}